\documentclass[useAMS,usenatbib,usegraphicx]{mn2e}
\usepackage{epsf}
\usepackage{times}
\usepackage{longtable}
\usepackage{fixltx2e} 

\usepackage[usenames]{color}
\definecolor{Blue}{rgb}{0.1,0,0.55} 
\definecolor{Red}{rgb}{1.0,0.1,0.1} 
\usepackage{hyperref}
\usepackage{breakurl}

\definecolor{ctcolor}{rgb}{0,0.42,0}
 
\newcommand{\tablefootmark}[1]{$^{#1}$}
\newcommand{\tablefoottext}[1]{$^{#1}$~}

\title[Bandmerged \textit{Planck} ERCSC]{The bandmerged \textit{Planck} Early Release Compact Source Catalogue: Probing sub-structure in the molecular gas at high Galactic latitude}

\author[X. Chen et al.]{
X.~Chen,$^{1}$
R.~Chary,$^{1,3}$
T.~J.~Pearson,$^{1, 2}$
P.~McGehee,$^{1}$
J.~W.~Fowler$^{1}$
and
G.~Helou$^{1}$\\
$^{1}$Infrared Processing and Analysis Center, California Institute of Technology, Pasadena, CA 91125, U.S.A.; \rm{Email: xchen@ipac.caltech.edu, rchary@caltech.edu}\\
$^{2}$Cahill Center for Astronomy \& Astrophysics, California Institute of Technology, Pasadena, CA 91125, U.S.A.\\
$^{3}$Smithsonian Astrophysical Observatory, Harvard-Smithsonian Center for Astrophysics, 60 Garden St, Cambridge, MA 02138, U.S.A.}

\begin{document}

\maketitle

\begin{abstract}
The \textit{Planck} Early Release Compact Source Catalogue (ERCSC) includes nine lists of highly reliable sources, individually extracted at each of the nine \textit{Planck} frequency channels. To facilitate the study of the \textit{Planck} sources, especially their spectral behaviour across the radio/infrared frequencies, we provide a  ``bandmerged'' catalogue of the ERCSC sources. This catalogue consists of 15191 entries, with 79 sources detected in all nine frequency channels of \textit{Planck} and 6818 sources detected in only one channel. We describe the bandmerging algorithm, including the various steps used to disentangle sources in confused regions. The multi-frequency matching allows us to develop spectral energy distributions of sources between 30 and 857\,GHz, in particular across the 100\,GHz band, where the energetically important CO $J$=1$\rightarrow$0 line enters the \textit{Planck} bandpass. We find $\sim3-5\sigma$ evidence
for contribution to the 100 GHz intensity from foreground CO along the line of sight to 147 sources with $|b|>30\degr$. The median excess contribution is 4.5$\pm$0.9\% of their
measured 100 GHz flux density which cannot be explained by calibration or beam uncertainties. This translates to 0.5$\pm$0.1\,K\,km\,s$^{-1}$ of CO which must be clumped on the scale of the
\textit{Planck} 100 GHz beam, i.e., $\sim$10$\arcmin$. If this is due to a population of low mass ($\sim$15\,M$_{\sun}$) molecular gas clumps, the total mass in these clumps 
may be more than 2000\,M$_{\sun}$.
Further, high-spatial-resolution, ground-based observations of the high-latitude sky will help shed light on the origin of this diffuse, clumpy CO emission.
\end{abstract}

%%%%%%%%%%%%%%%%%%%%%%%%%%%%%%%%%%%%%%%%%%%%%%%%%%%%%%%%%%%%%%%%%%%%%%%%%%%%%%%%%
\section{Introduction}
\label{sec:introduction}

From 2009 August to 2012 January, the \textit{Planck}\footnote{\textit{Planck}
(http://www.esa.int/Planck) is a project of the European Space Agency (ESA) with instruments provided by two scientific consortia funded by ESA member states (in particular the lead countries France and Italy), with contributions from NASA (USA) and telescope reflectors provided by a collaboration between ESA and a scientific consortium led and funded by Denmark.} satellite carried out high sensitivity all-sky surveys at nine frequencies from 30 to 857\,GHz, with beam sizes ranging from about 33 to 4.2\,arcmin \citep{planck2011-1.1}. The \textit{Planck} Early Release Compact Source Catalogue \citep{planck2011-1.10} lists sources detected in the first 1.6 full-sky survey maps, and offers a unique opportunity for studying these sources, both Galactic and extragalactic, in a previously under-explored frequency domain. The ERCSC was released on 2011 January 11, and is available from both the ESA \textit{Planck} Legacy Archive\footnote{http://www.cosmos.esa.int/web/planck/pla} and the NASA Planck Archive.\footnote{http://irsa.ipac.caltech.edu/Missions/planck.html}

In the ERCSC, source detection and characterization were done independently in each \textit{Planck} band yielding source lists at each frequency. The minimum S/N in the compact
source kernel filtered individual frequency maps is 5. In addition, since the primary goal of the ERCSC was $>90$\% reliability, a Monte-Carlo algorithm based on the injection and extraction of
artificial sources into the Planck maps was implemented to select
reliable sources among all extracted 5$\sigma$ candidates such that the
cumulative reliability of the catalogue is $>$90\%. The reliability was
defined as the signal to noise where that the extracted flux density of the extracted artificial sources would agree with the
input flux density to better than 30\%. Thus, if a source fell on top
of a bright cirrus feature such that its flux density was biased by more than
30\%, it would be ignored. Since Galactic cirrus emission is the
dominant contaminant at the higher frequencies, and cirrus emission is
patchy, the completeness at a particular flux density can be high in
clean patches of sky but low in regions where the ISM emission is
strong. The reliability cuts are applied per band
with a full description of the process provided in \citet{planck2011-1.10}.
Here, we merged the catalogs per band after these cuts;
so whether
or not a source is a multiband source is not known until after the
bandmerging process.

While the source lists in individual bands are important in characterizing the statistical properties of source populations at each frequency \citep{planck2011-6.1, planck2012-VII}, the multi-frequency power of \textit{Planck} can only be fully exploited when the sources detected in each band are associated. 
A suite of \textit{Planck} Early Results papers have already shown the power of multifrequency SED in finding gigahertz peaked spectrum (GPS) sources \citep{planck2011-6.2}, identifying anomalous microwave emission regions and associating the emission with spinning dust \citep{planck2011-7.2}, and studying the synchrotron break frequencies of blazars \citep{planck2011-6.3a}. To better facilitate the study of spectral properties of the \textit{Planck} sources by the scientific community, we have attempted to find associations between  the sources detected at different frequencies in the ERCSC, a process known as ``bandmerging.'' 

In Section~\ref{sec:algorithm} we describe the bandmerge algorithm, with emphasis on confusion processing. In Section~\ref{sec:bmcat} we present  the bandmerged catalogue and discuss the spectral energy distributions of the sources. In Section~\ref{sec:co} we use the bandmerged catalogue to estimate the flux-density excess due to CO contamination in the \textit{Planck} 100\,GHz channel and use this information to assess the CO intensity at high Galactic latitudes. Section~\ref{sec:conclusion} presents our conclusions.

%%%%%%%%%%%%%%%%%%%%%%%%%%%%%%%%%%%%%%%%%%%%%%%%%%%%%%%%%%%%%%%%%%%%%%%%%%%%%%%%%

\section{Bandmerge Algorithm}
\label{sec:algorithm}

Following \citet{Fowler2003},
%the technical document ``Bandmerging \textit{Spitzer} Point Sources'' by J. W. Fowler, 
we designed an algorithm that suits the specific needs of the ERCSC. The main challenges we face in the merging process are: (1) different resolution across the bands, which leads to source confusion; and (2) the broad frequency coverage, which induces complications due to the interstellar medium (ISM) emission and the transition from frequencies where the synchrotron and free-free emission dominate to where the thermal dust emission becomes significant. 

\subsection{Cross-band matching}
\label{sec:crossband}
We begin by selecting a source in one \textit{Planck} band (the ``seed'' band), and searching all the other eight bands (the ``candidate'' bands) for counterparts. The matching radius is always chosen to be half of the full width at half maximum (FWHM) of the larger of the two beams between the seed and candidate bands. Table~\ref{tab:basics} provides the frequency and beam width of each of the \textit{Planck} bands.  {\it Planck} is not a diffraction-limited telescope and the resolution is limited by the scan rate, the sparse focal plane and under-illumination of the primary; therefore the resolution at
the upper frequencies is nearly the same.
Since the upper four bands have very similar angular resolutions, we adopted 5\,arcmin as the uniform beam size for these four bands in the matching process. We link a source in the candidate band to the seed source when the positional offset between them is less than the matching radius. The positional accuracy of the ERCSC sources is estimated to be better than 1/5 of the FWHM at each band \citep{planck2011-1.10}. Therefore our choice of the matching radius reflects a balance between including the correct counterparts of our seed source in the candidate bands, and excluding close but unrelated neighbours at the same time. To avoid biasing towards or against any class of astrophysical sources, we do not use any brightness information at this step. Whether a detection in the candidate band is an acceptable match to the seed source is decided exclusively on the basis of position. 

\begin{table*}
\begin{minipage}{110mm}
\caption{Characteristics of the \textit{Planck} bands\label{tab:basics}}
\begin{tabular}{lccccccccccc}
\hline
 & \multicolumn{3}{c}{LFI} &  & & \multicolumn{6}{c}{HFI} \\
\hline
\textit{Planck} Band [GHz] & 30 & 44 & 70 & & & 100 & 143 & 217 & 353 & 545 & 857 \\
Band Pass [$\Delta \nu$, GHz] & 6 & 9 & 14 & & & 32.9 & 45.8 & 64.5 & 101.4 & 171.3 & 245.9 \\
Beam FWHM [arcmin] & 32.65 & 27.00 & 13.01 & & & 9.94 & 7.04 & 4.66 & 4.41 & 4.47 & 4.23 \\
\quad\quad(adopted value$^{a}$) &  &   &  & & &   &  & 5.00 & 5.00 & 5.00 & 5.00 \\
\hline
\end{tabular}

\medskip
$^{a}$ The upper four bands have similar angular resolutions, we therefore adopted 5\,arcmin as the uniform beam size for these four bands in the bandmerging process.
\end{minipage}
\end{table*}

Each of the nine \textit{Planck} bands is used as seed band once. For each source in the seed band, we save the total number of acceptable matches (i.e., those with positional offsets less than the matching radius) in each candidate band to quantify the seriousness of the confusion at that frequency. However, we only save pointers to up to three qualifying matches per candidate band, arranged in ascending order of the distance to the seed source. This choice was somewhat arbitrary, although we later found that the maximum number of acceptable matches per candidate band for each seed source in our analysis is actually three.  

\subsection{Confusion Processing}
\label{sec:confusion}
Ideally, after the cross-band matching, for each given source in a band, there would be at most one pointer to a counterpart in another band, with the counterpart having only one pointer pointing back to the given source. In addition, every source in a single merged chain should point to the others. This is indeed the most common case. However, in regions of high source density (typically at low Galactic latitudes), sources that are separated by less than the matching radius often get erroneously cross-associated with each other. We describe below the four different types of confusion cases we encountered in the ERCSC bandmerging process and our solutions. 

\subsubsection{Inconsistent chain}

\begin{figure}
\centering
\includegraphics[width=0.44\textwidth]{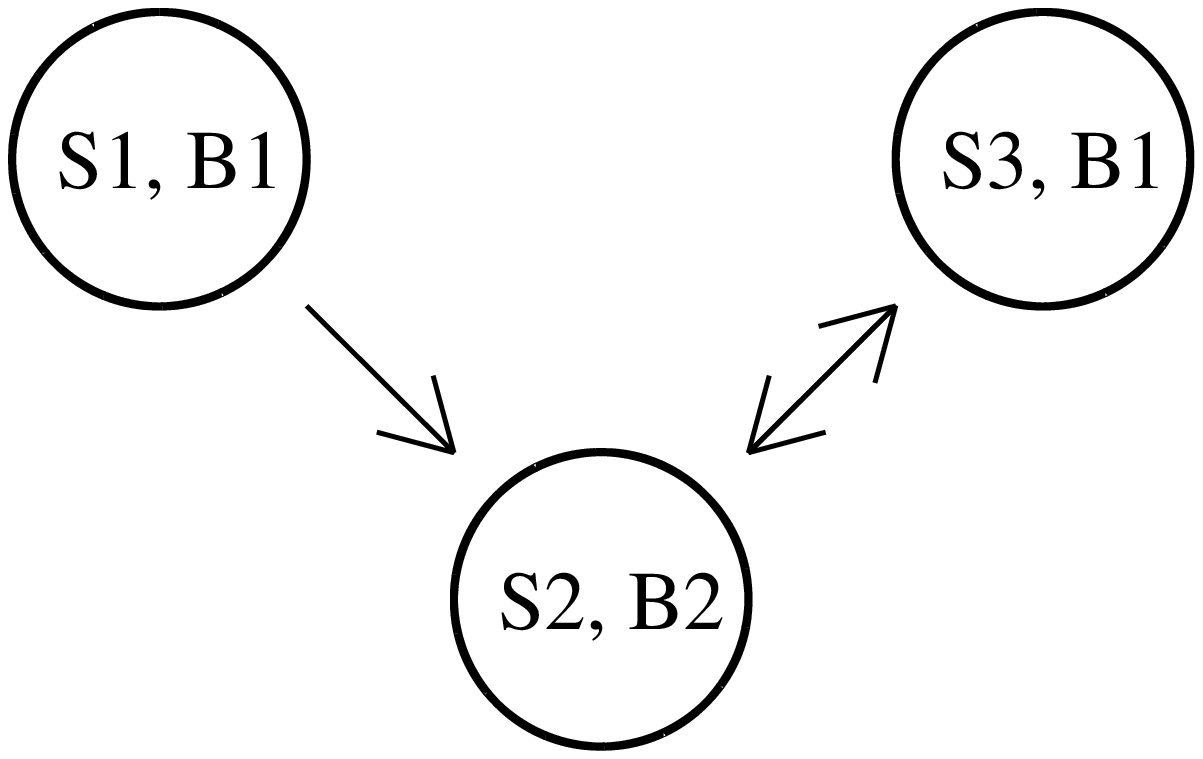}
\includegraphics[width=0.37\textwidth]{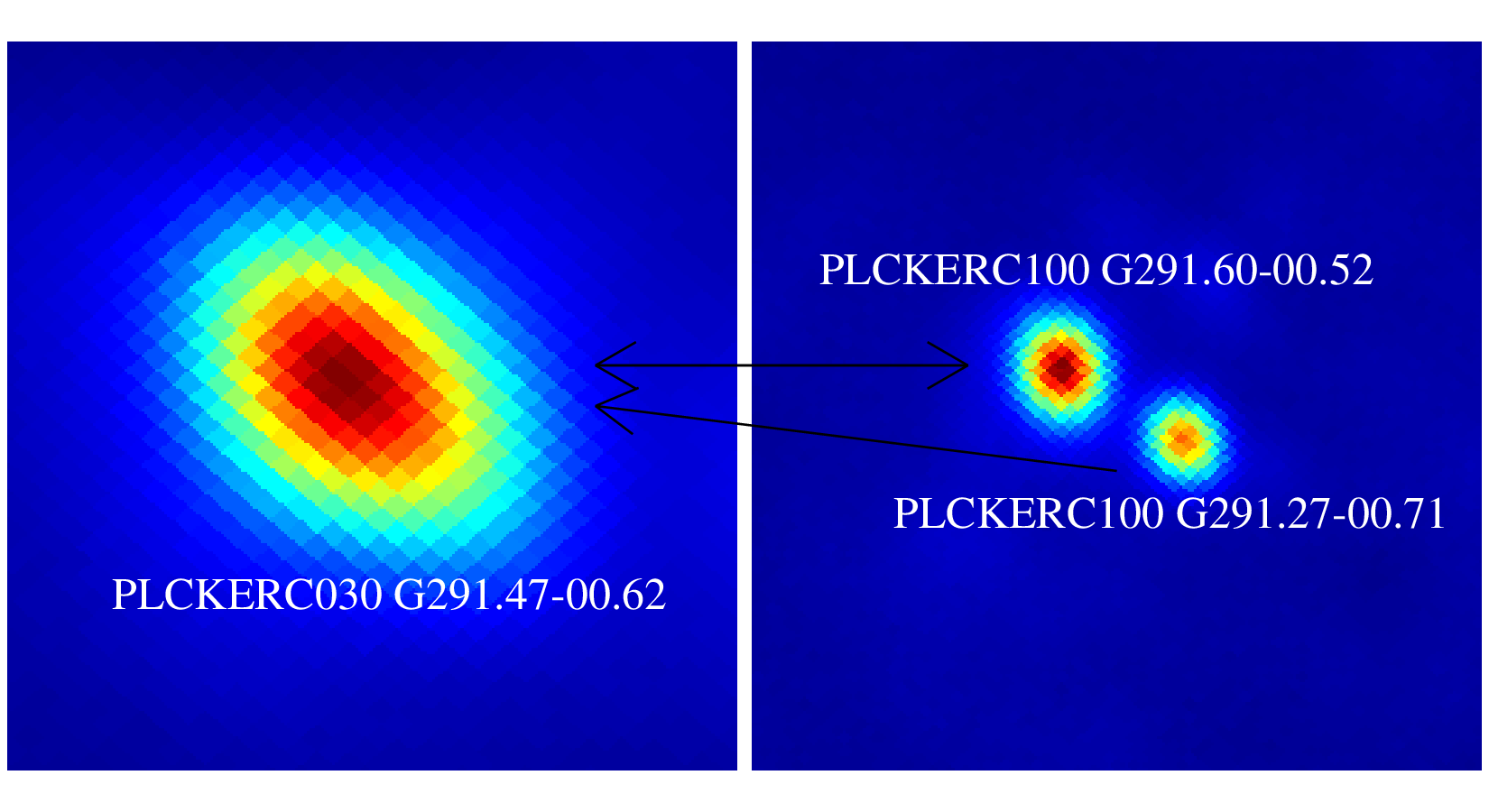} 
\caption{Inconsistent chain model (top) and real example in bandmerging process (bottom). One-direction arrow goes from seed source to preferred candidate source, while dual-direction arrow indicates reciprocal relation between seed and candidate. The closest 30\,GHz match for PLCKERC100 G291.27$-$00.71 is PLCKERC030 G291.47$-$00.62. However, when searching the source list at the 100\,GHz band, we find PLCKERC100 G291.60$-$00.52 is the closest match for this 30\,GHz source. 
\label{fig:inconsistent_chain}
}
\end{figure}

``Inconsistent chain'' refers to the situation where a given seed source's first-choice counterpart in a candidate band does not have a reciprocal first-choice pointer back to the given source. A model of an inconsistent chain is given in Figure~\ref{fig:inconsistent_chain}, where source S1 in band B1 finds source S2 as the preferred match (i.e., closest neighbour) in band B2.  However, when source S2 is used as the seed source, it finds source S3 in band B1 is a better match, i.e., closer to source S2. In the lower part of Figure~\ref{fig:inconsistent_chain}, we show a real example of inconsistent chain in our bandmerging process. When we use 100\,GHz as the seed band and 30\,GHz as the candidate band, source PLCKERC100 G291.27$-$00.71 finds PLCKERC030 G291.47$-$00.62 as the closest match. However,  when the 30\,GHz band is used as seed band, the same 30\,GHz source finds PLCKERC100 G291.60$-$00.52 as the closest match in the 100\,GHz band.  

To solve inconsistent chains, we loop over all source records, and check that the preferred pointers are symmetric under interchange of seed and candidate sources. If not, we follow the chain of pointers until a reciprocal relation is found, and break the previous link. Whenever a linkage is broken, any existing second-choice pointer is elevated to first-choice status, and any existing third-choice pointer is elevated to second-choice status. We then begin processing the current source records from the beginning until only reciprocal links remain. In our example, PLCKERC100 G291.60$-$00.52 and PLCKERC030 G291.47$-$00.62 are found to be the preferred matches to each other.  We therefore keep the link between them and break the one between the 30\,GHz source and PLCKERC100 G291.27$-$00.71. 

Due to resolution differences between the two bands, the 30\,GHz source could also be a blend of the two 100\,GHz sources. We therefore further check the flux densities and inspect multifrequency cutout images around our sources of interest using the ERCSC visualization tool at the \textit{Planck} IRSA archive. In this case, the images reveal that PLCKERC100 G291.27$-$00.71 is a dusty source that only starts to emerge at 70\,GHz, and gets brighter towards the higher frequencies. Based on its 70--353\,GHz flux densities,  the contribution of this source to the 30\,GHz flux density could not be more than $25\%$ of the flux density of PLCKERC030 G291.47$-$00.62.

\subsubsection{Excess linkage}
\begin{figure}
\centering
\includegraphics[width=0.44\textwidth]{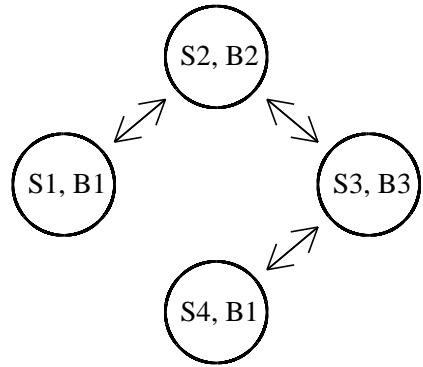}
\includegraphics[width=0.37\textwidth]{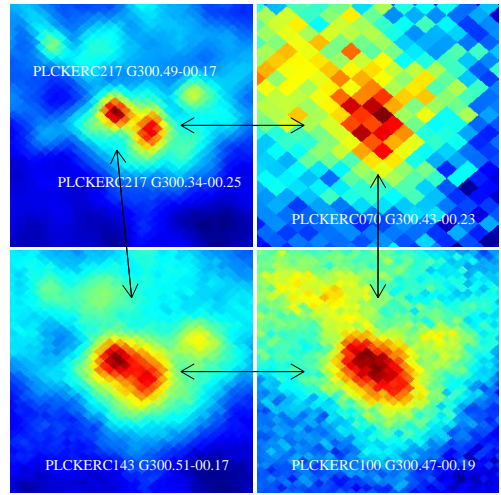} 
\caption{Excess linkage model (top) and real example in bandmerging process (bottom). Dual-direction arrow indicates reciprocal relation between seed and candidate. In the real example here, if we start with PLCKERC217 G300.34$-$00.25, all the reciprocal relations will lead us to PLCKERC143 G300.51$-$00.17. However, this 143\,GHz source has a reciprocal relationship with a different source PLCKERC217 G300.34$-$00.25 at the 217\,GHz band. \label{fig:excess_linkage}
}
\end{figure}

After all the linkages have been made reciprocal, inconsistencies can still exist. For example (see Figure~\ref{fig:excess_linkage}), a detection S1 in band B1 may have a reciprocal relationship with a detection S2 in band B2, which has a reciprocal relationship with a detection S3 in band B3, which then has a reciprocal relationship with a different source S4 in band B1 than the detection that started the chain. We therefore have two sources in band B1 in the same merged chain. Such situation is referred to as excess linkage. In the ERCSC bandmerging process, excess linkage often happens when one source detected in a lower-resolution band is resolved into multiple sources in higher-resolution bands. We break the links between the unresolved detections and the resolved ones, and separate them as different entries in the bandmerged catalogue. In the real case we present in Figure~\ref{fig:excess_linkage}, a hotspot that is seen as one source from 70 to 143\,GHz ($\sim 7$\,arcmin resolution) is resolved into two distinct sources at 217\,GHz ($\sim 5$\,arcmin resolution). In this case, PLCKERC070 G300.43$-$00.23, PLCKERC100 G300.47$-$00.19 and PLCKERC143 G300.51$-$00.17 would remain linked and be listed as one entry in the merged catalogue, with a note indicating its association with the two sources at 217\,GHz that are listed separately as two independent entries.

\subsubsection{Linkage rejection}
\begin{figure}
\centering
\includegraphics[width=0.44\textwidth]{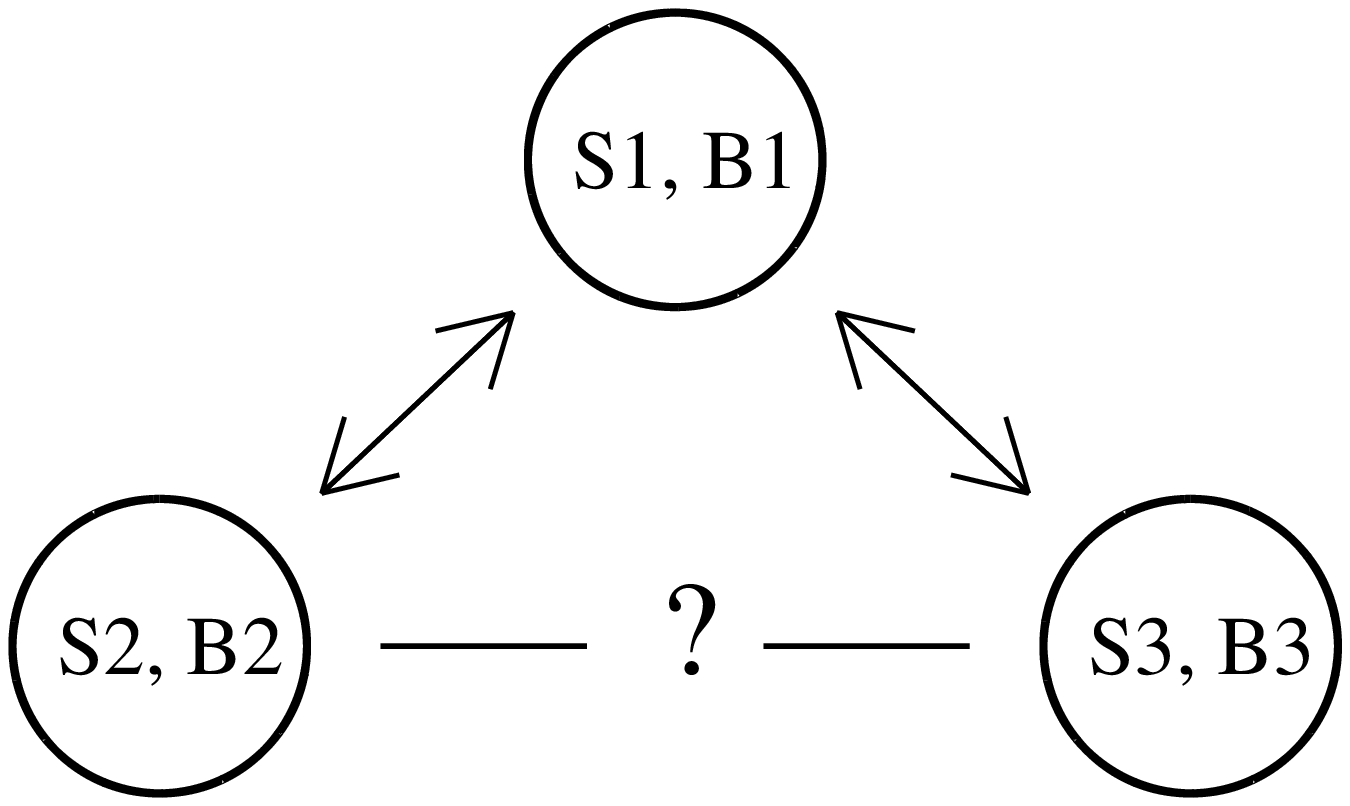}
\includegraphics[width=0.37\textwidth]{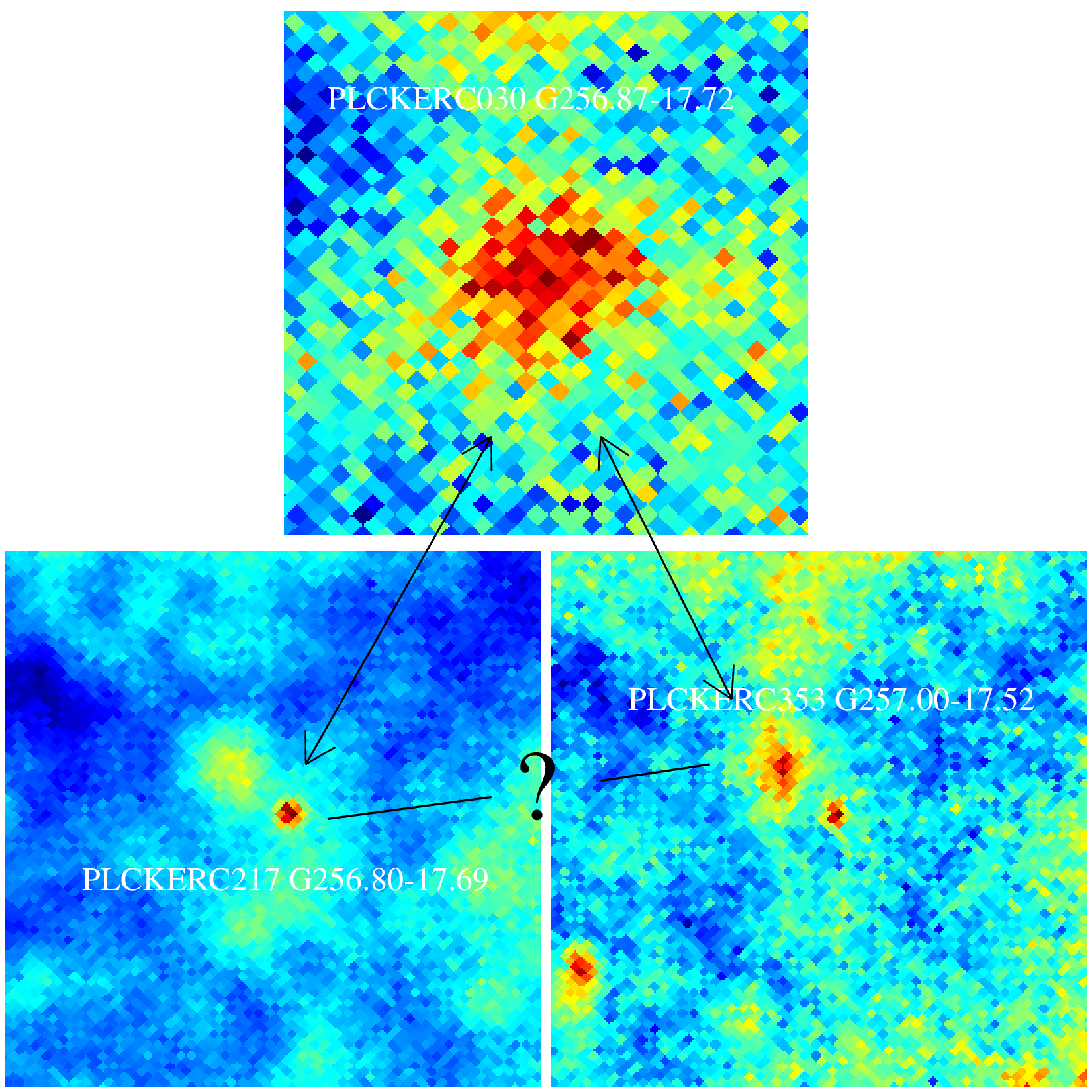} 
\caption{Linkage rejection model (top) and real example in bandmerging process (bottom). Dual-direction arrow indicates reciprocal relation between seed and candidate, while short broken lines indicates mutually-rejected sources. The 30\,GHz source PLCKERC030 G256.87$-$17.72 is seen resolved into two components at both 217 and 353\,GHz: PLCKERC217 G256.80$-$17.69 is detected at 217\,GHz but is not in the 353\,GHz catalogue, while PLCKERC353 G257.00$-$17.52 is detected at 353\,GHz but is not in the 217\,GHz catalogue. The extended nature suggests that the latter may be thermal dust emission from the ISM. \label{fig:linkage_rejection}
}
\end{figure}

Linkage rejection refers to the problem that not all detections in a single merged chain are linked to each other. We illustrate this problem using the model given in Figure~\ref{fig:linkage_rejection}, where source S1 in band B1 is matched to both source S2 in band B2 and source S3 in band B3, but S2 and S3 are not matched to each other. For bands with similar resolutions, this problem is usually caused by the different positional errors of the detections in different bands. In the case of ERCSC, the main cause is the difference in the beam sizes in different band, which means that different matching radii are used for each pair of bands during the process of cross-band matching. For instance, assuming source S1 is from the 30\,GHz band, sources S2 and S3 are from 217\,GHz and 353\,GHz bands individually. As defined in Section~\ref{sec:crossband}, the matching radii between the 30 and 217\,GHz bands and the 30 and 353\,GHz bands are both $\sim 16$\,arcmin as the 30\,GHz band has the larger beam in both matching pairs. Therefore sources S2 and S3 can both be associated with source S1 if the positional offsets of these two pairs are both less than 16\,arcmin. However, when source S2 and source S3 are compared, the matching radius is 0.5 FWHM of the 217 or 353\,GHz beam, 2.5\,arcmin. Therefore, these two sources will not be matched  if they are further than 2.5\,arcmin apart. However, the fact that S2 and S3 are indirectly linked to a common source gives some credibility that they are probably matches to each other. We therefore test the unlinked detections with an enlarged matching radius of 1 FWHM. If this relaxed threshold is passed, these detections are temporarily associated. We then use the ERCSC visualization tool to inspect the multifrequency images of these sources. If the images confirm that the detections at different frequencies are indeed the same source, we make the linkage permanent. If the images suggest that the detection made at the lower-resolution band is the blending of the detections at higher-resolution bands, we break the linkage. Figure \ref{fig:linkage_rejection} presents such an example, where the 30\,GHz detection PLCKERC030 G256.87$-$17.72 breaks into two components at higher frequencies. The more compact component in the center is detected at 217\,GHz, but is not in the 353\,GHz catalogue; while the component to the upper left of the center is detected at 353\,GHz but not 217\,GHz. Although the compact 217\,GHz source is seen at 353\,GHz (Figure~\ref{fig:linkage_rejection}), its high frequency counterpart did not meet the SNR requirements of the ERCSC and was excluded from the 353\,GHz catalogue. We link PLCKERC217 G256.80$-$17.69 and  PLCKERC353 G257.00$-$17.52 temporarily after they survive the larger matching radius, but eventually break the links between these two sources as it is evident from the images that they are two distinct sources. We further examine the SED of the sources and find that PLCKERC217 G256.80$-$17.69 is the correct counterpart of PLCKERC030 G256.87$-$17.72. We therefore keep the link between these two and merge them into one entry in the bandmerged catalogue.

\subsubsection{Cirrus contamination}

\begin{figure}
\centering
\includegraphics[width=0.37\textwidth]{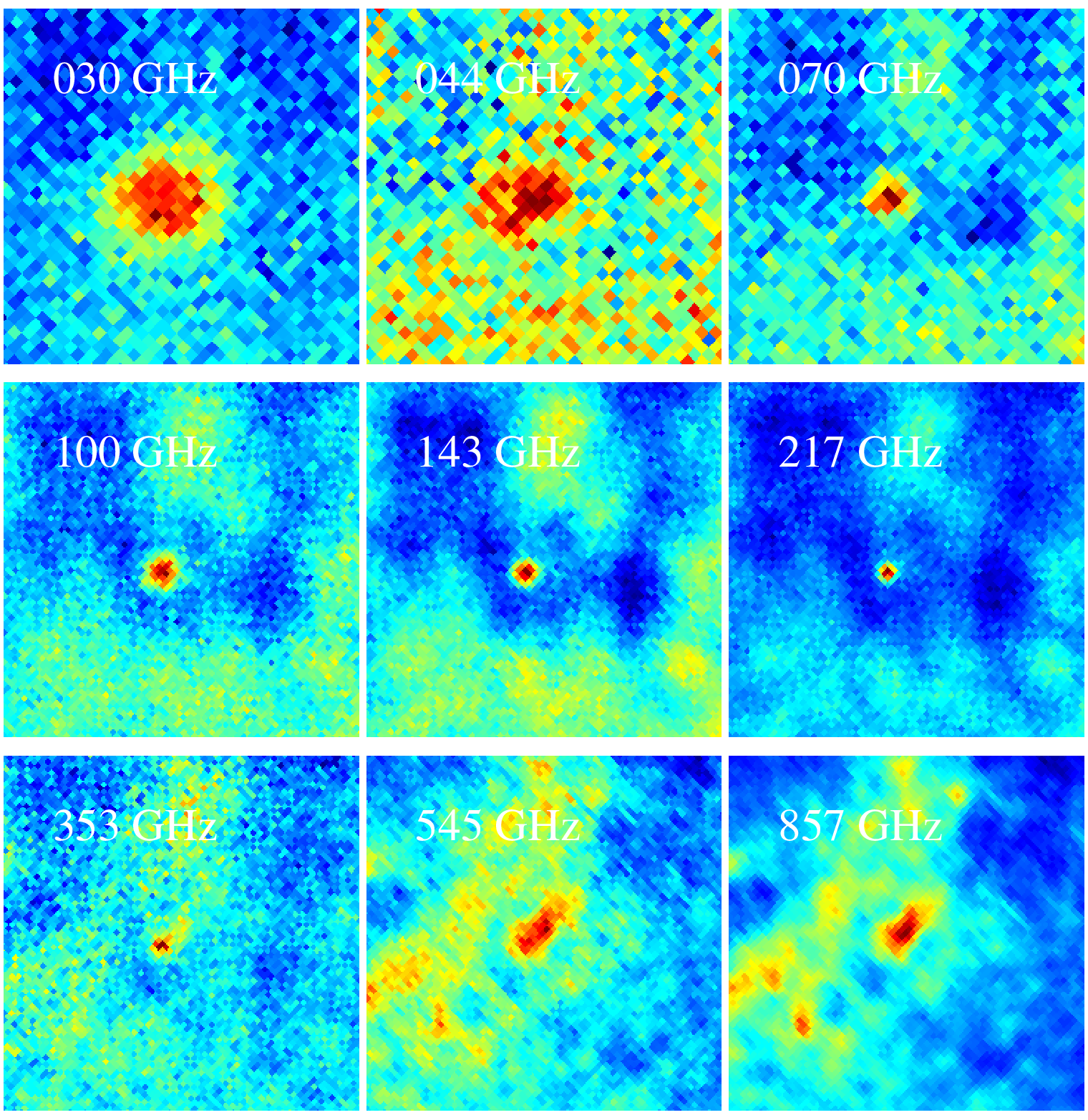}
\includegraphics[width=0.41\textwidth]{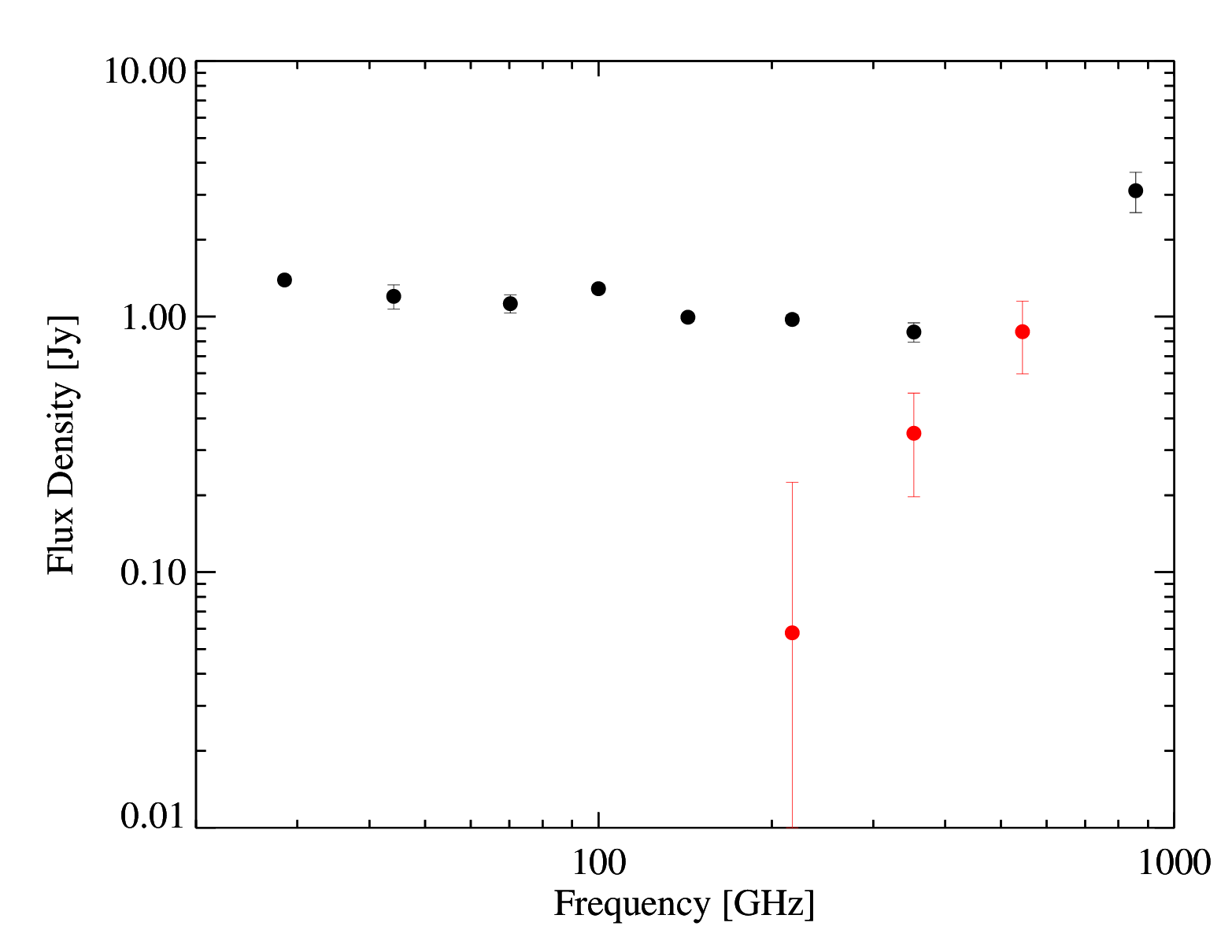} 
\caption{Example of a compact blazar (4C +56.27) misassociated with extended ISM emission during the bandmerging process. Top: 2 by 2 degree images centered on the source in \textit{Planck} 30--857\,GHz maps. Bottom: SED constructed from the merged detections (black filled circles), and from the bandfilled flux densities of the 857\,GHz source at 217, 353 and 545\,GHz (red filled circles). In some cases, the uncertainty in the flux density is smaller than the size of the symbol and so the error bar is not shown. The clean flat spectrum between 30 and 353\,GHz suggests that the association in these frequency channels is solid. The discrepancies seen between the bandfilled flux densities of the 857\,GHz source and the flat spectrum indicate that the 857\,GHz detection is slightly offset from the quasar, and its 857\,GHz flux density is mostly from the ISM emission. \label{fig:cirrus}}
\end{figure}

The aforementioned steps result in a bandmerged catalogue with reciprocal one-to-one pointers between matched sources, and in which all the sources linked in each merged chain are pointed to each other. However, between the 353 and 857\,GHz frequency channels, the contribution from the ISM emission gets stronger and Galactic cirrus features can sometimes be confused with compact sources. An example is given in the top panel of Figure \ref{fig:cirrus}, where a detection is made at the center of the field in all the \textit{Planck} bands except for the 545\,GHz band. All the detections are linked to each other in a one-to-one reciprocal relationship in a single merged chain. The 30--353\,GHz images show a bright, compact source in the center, and a positional search in the NASA/IPAC Extragalactic Database\footnote{\url{http://ned.ipac.caltech.edu/}} confirms the source as the well-known blazar 4C +56.27. The 545 and 857\,GHz images, however, clearly reveal the structure of extended ISM emission which physically overlaps the blazar detected at the lower frequencies. Therefore, we cannot blindly associate the blazar with the detection at 857\,GHz since the ISM emission at 857\,GHz would boost the flux density of the blazar at the resolution of \textit{Planck} and result in a rising spectrum which may not be truly associated with the blazar. We plot the SED of this merged chain in the bottom panel of Figure \ref{fig:cirrus}. The black filled circles are the flux densities of all the detections in the merged chain. The clean, flat spectrum of the blazar between 30 and 353\,GHz suggests the association at these frequency channels is robust. The jump at 857\,GHz is due to the contamination by ISM emission. We overplot the bandfilled flux densities for the 857\,GHz sources\footnote{For each source detected in the 857\,GHz band, the ERCSC also had entries which contained the flux densities of the source in the 217, 353 and 545\,GHz maps derived through photometry in apertures centered on the 857\,GHz position of the source. These flux densities are referred to as bandfilled flux densities, and are included in the ERCSC 857\,GHz source list.} at 217, 353, and 545\,GHz in red. Large discrepancies can be seen between the bandfilled flux densities of the 857\,GHz detection and the flat spectrum of the blazar at 217 and 353\,GHz. This suggests that there is a small offset between the 857\,GHz source position (the peak of the cirrus feature) and the position of the blazar. The bandfilled flux densities and the 857\,GHz flux density constitute a smooth rising spectrum that is a signature of thermal dust emission, which further confirms that the detection at 857\,GHz is mostly an ISM feature. It is also noticeable here that the 100\,GHz flux density of the quasar is a little high. This is due to the contamination from the CO $J$=1$\rightarrow$0 emission line in the 100\,GHz bandpass of \textit{Planck}, which we will discuss in more detail in Sections~\ref{sec:all9} and~\ref{sec:co}. 

We detect and disassociate cirrus contamination by first selecting sources with bandmerged 217--545\,GHz flux densities that differ from the bandfilled flux densities at these frequencies by more than the total of uncertainties from both the bandmerged and the bandfilled flux densities. Since the ISM emission is normally extended -- as in the example shown here -- and the upper four bands have very similar spatial resolution, if the sources detected at 217 and 353\,GHz are compact, then the 545 and 857\,GHz counterparts should also be compact for robust matches.
The EXTENDED flag in the ERCSC is used to selected sources which are extended at 857 or 545\,GHz
but compact at 217 or 353\,GHz and these are identified as cirrus candidates. Finally we inspected the images and SEDs of these cirrus candidates, keeping the robust matches and separating the incorrect associations.

%%%%%%%%%%%%%%%%%%%%%%%%%%%%%%%%%%%%%%%%%%%%%%%%%%%%%%%%%%%%%%%%%%%%%%%%%%%%%%%%%
\section{The Bandmerged Catalogue}
\label{sec:bmcat}

\subsection{Content}

\begin{table*}
\begin{minipage}{100mm}
\caption{Bandmerged ERCSC content\label{tab:bmcat}}
\begin{tabular}{ll}
\hline
Parameter & Description \\
\hline
NAME & Source name  \\
GLON & Galactic longitude [deg]  \\
GLAT & Galactic latitude [deg]  \\
RA & Right ascension in J2000 [deg]  \\
DEC & Declination in J2000 [deg]  \\
NBAND & Number of \textit{Planck} bands with detection \\
FLUX & Flux density [mJy] in the \textit{Planck} bands, set to NaN if not available. \\ 
FLUX\_ERR & Flux density uncertainty [mJy] in the \textit{Planck} bands, set to NaN if not available. \\
ERCSC\_NAME & Name of the source in the ERCSC, set to blank if not available.  \\
NOTE & Flag, set to 1 if there is an entry in the notes file. \\
\hline
\end{tabular}
\end{minipage}
\end{table*}

The final bandmerged ERCSC has 15191 entries. It is distributed on the form of a FITS binary table, and is available in the NASA \textit{Planck} Archive. The columns are described in Table \ref{tab:bmcat}. Sources are sorted in ascending order of Galactic longitude. The entry name and positions are based on the detection at the highest frequency in a merged chain, as it is generally true for \textit{Planck} that the higher the frequency, the better the angular resolution and the better the positional accuracy of the source. 
The spatial resolution for each of the upper four bands is comparable, therefore it does not matter which of the upper frequency counterparts we base the position on; we use the position from the detection at the highest frequency for consistency with the choice at lower frequencies. For instance, the first entry of the catalogue is PLCKERC G000.01$-$18.90. The highest frequency at which this source is detected is 857\,GHz.\, so its NAME is inherited from the 857\,GHz counterpart PLCKERC857 G000.01$-$18.90 (with the frequency omitted), and its GLON, GLAT, RA and DEC are propagated from the 857\,GHz source. This choice of the position for the merged entry may not always be the best solution, as the emission centroid of a source can sometimes shift between radio and infrared wavelengths. We therefore encourage the users to refer to the ERCSC single-band source lists for characteristics of a source derived in each frequency channel.

The number of frequency channels (1--9) in which a source is detected is recorded in the NBAND column. FLUX and FLUX\_ERR are both 9-element vectors that are set to the ERCSC FLUX and FLUX\_ERR \citep{planck2011-1.10} in the bands where a detection is made, and set to NaN in bands where there is no ERCSC detection. Similarly, the names of the sources in the ERCSC single-band lists are provided in ERCSC\_NAME, with a null value indicating no detection at a given band. We distribute with the merged catalogue a note file, providing additional notes for sources with NOTE flag set to~1.

\subsection{Spectral types}
\begin{figure*}
\centering
\includegraphics[width=0.45\textwidth]{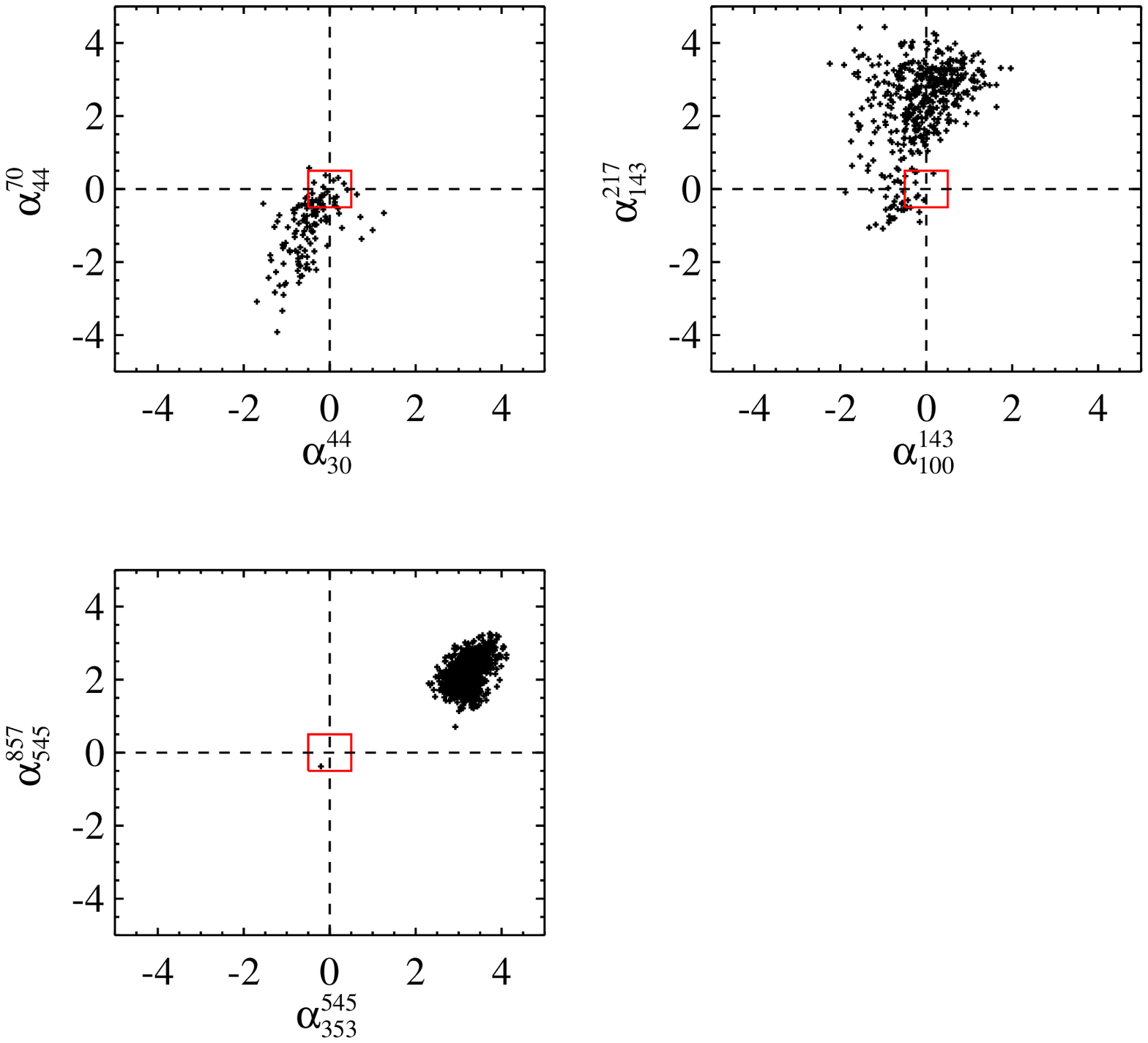}
\includegraphics[width=0.45\textwidth]{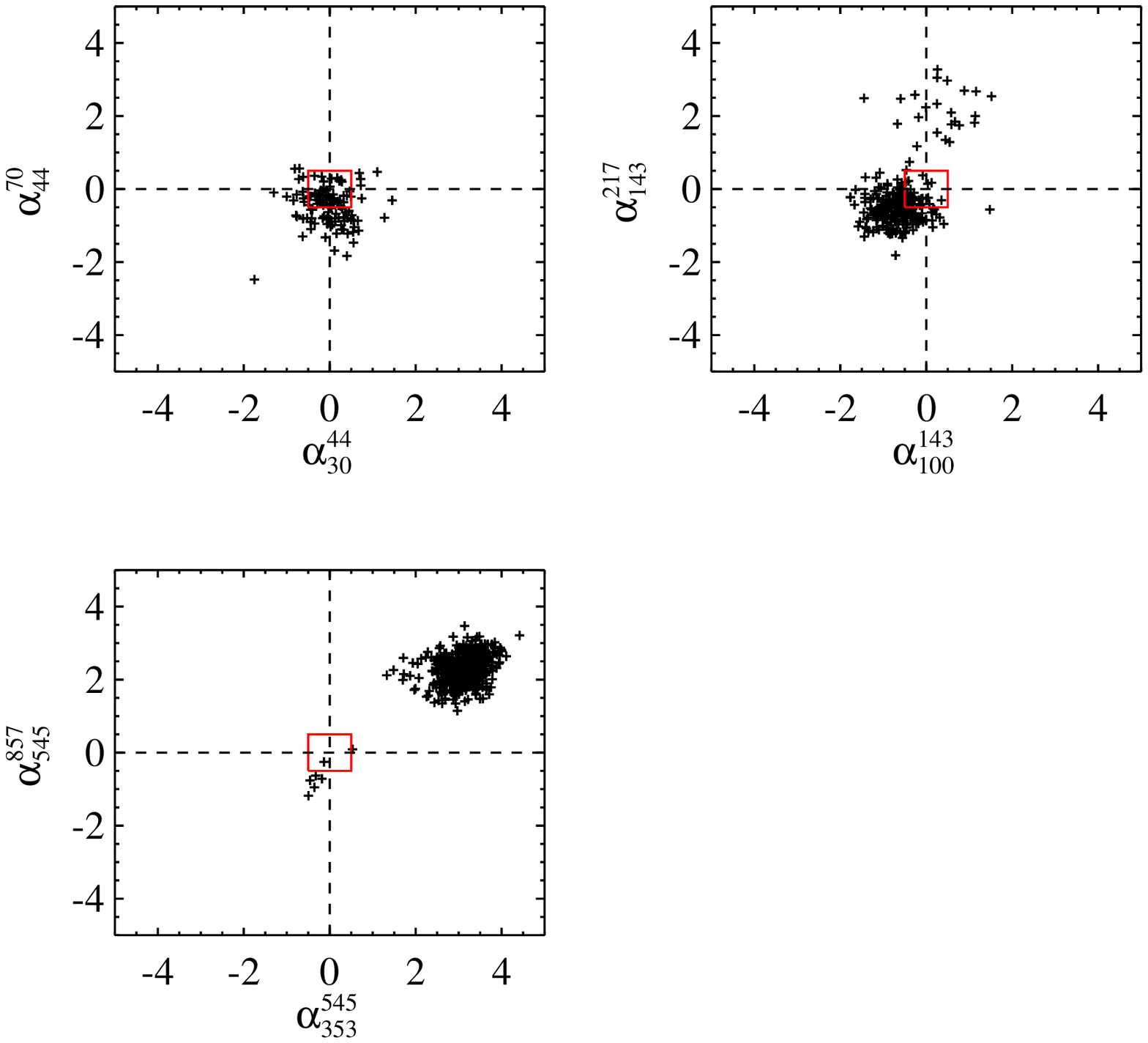} 
\caption{Colour-colour plot for sources at $|b| < 10\degr$ (left) and $|b| > 30\degr$ (right). The upper left quadrant corresponds to sources with upturn spectra, the upper right quadrant to rising-spectrum sources, the lower right quadrant to peaked-spectrum sources, and the lower left quadrant to steep-spectrum sources. The red box in the centre defines sources with flat spectra.\label{fig:ccplot}}
\end{figure*}

Using the bandmerged catalogue, we have selected sources that are detected in all three of the LFI bands (319 in total), in the 100--217\,GHz bands (910 in total), and in the 353--857\,GHz bands (3376 in total). Within each group, we have calculated spectral indices $\alpha~(S_{\nu} \propto \nu^{\alpha})$ between the adjacent bands. Figure~\ref{fig:ccplot} presents the colour-colour diagrams for the spectral indices between 30, 44 and 70\,GHz; 100, 143 and 217\,GHz; and 353, 545 and 857\,GHz. Separate colour-colour diagrams are presented for sources at $|$b$| < 10\degr$ that are likely Galactic, and for sources at $|$b$| > 30\degr$ that are likely extragalactic. Although not all high latitude sources may be extragalactic, statistically, they are likely dominated by extragalactic sources and appear to show
a different distribution of spectral indices compared to sources close to the Galactic Plane.
As can be seen, at low latitudes, the low frequency sources are predominantly synchrotron dominated sources while at high latitude,
they have flat spectra, indicating the dominant role of radio galaxies and blazars. At intermediate frequencies, there is clearly a transition to free-free emission and thermal dust emission in the
Plane while at high latitude, the population is still dominated by blazars with a flat spectrum. At the highest frequencies, we see primarily thermal dust emission both at low and high Galactic latitudes.
Since the ERCSC is a high reliability, signal to noise limited catalogue, with varying flux density across the sky, there are a multitude of source classes which are sampled by the catalogue. 
With the large beam of {\it Planck} it is challenging to make associations with
counterparts at other wavelengths, particularly the optical/near-infrared. Thus, the best attempt at source classification comes from the spectral indices across the {\it Planck} bands themselves.
This helped identify sources such as planetary nebulae, stars and supernova remnants close to the Plane. At high latitudes, the low frequency sources are associated with AGN/radio galaxies while the high frequency sources are predominantly ISM features and star-forming galaxies. 
An attempt to classify 57 sources selected at 857 GHz at $|$b$|>20\degr$, found that 44\% of the sources were associated with the ISM. Stars, galaxies, pre-stellar cold clumps and star-forming regions accounting for 12\% each, thereby highlighting the heterogeneity of the catalog \citep{Johnson}.

We consider all sources with $|\alpha_{\nu_1}^{\nu_2}| < 0.5$ to have a flat spectrum, where $\nu_1$ and $\nu_2$ are the two adjacent frequencies.  We highlight this region with a red box in the colour-colour plot. Excluding this region, the upper left, upper right, lower right and lower left quadrants correspond to upturn, rising, peaked, and steep spectrum sources, respectively. It is clear that at the radio frequencies, Galactic sources are dominated by steep spectrum, while the extragalactic sources are dominated by flat spectrum. At the intermediate frequencies, while the Galactic sources are clearly transitioning from a relatively flat spectrum between 100 and 143\,GHz to a rising spectrum between 143 and 217\,GHz, the extragalactic sources mostly have steep or flat spectra. At the highest frequencies, rising-spectrum, dusty sources dominate both Galactic and extragalactic source populations.

\begin{figure*}
\centering
 \begin{tabular}{ccc}
\includegraphics[width=0.31\textwidth]{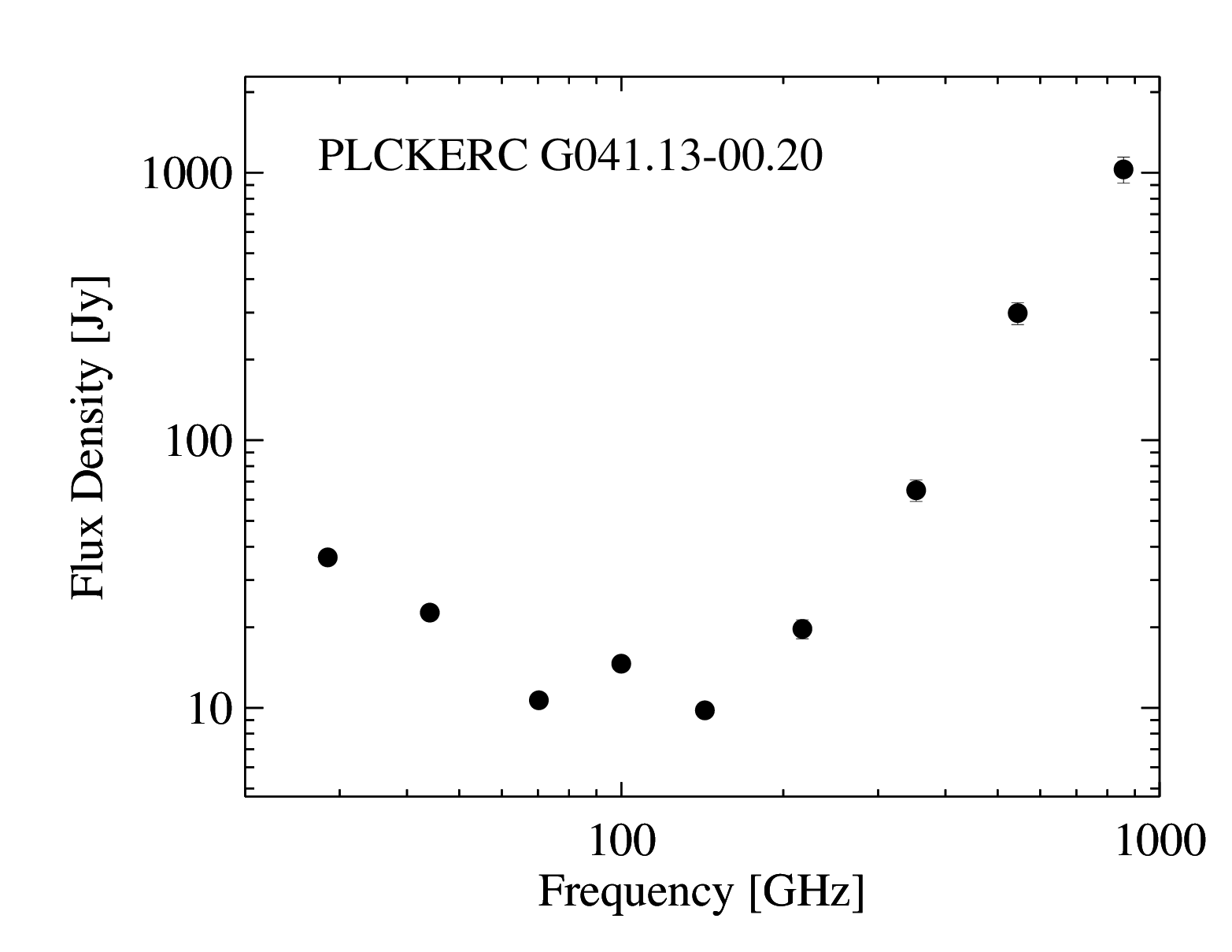} &  \hspace{-20pt}
\includegraphics[width=0.31\textwidth]{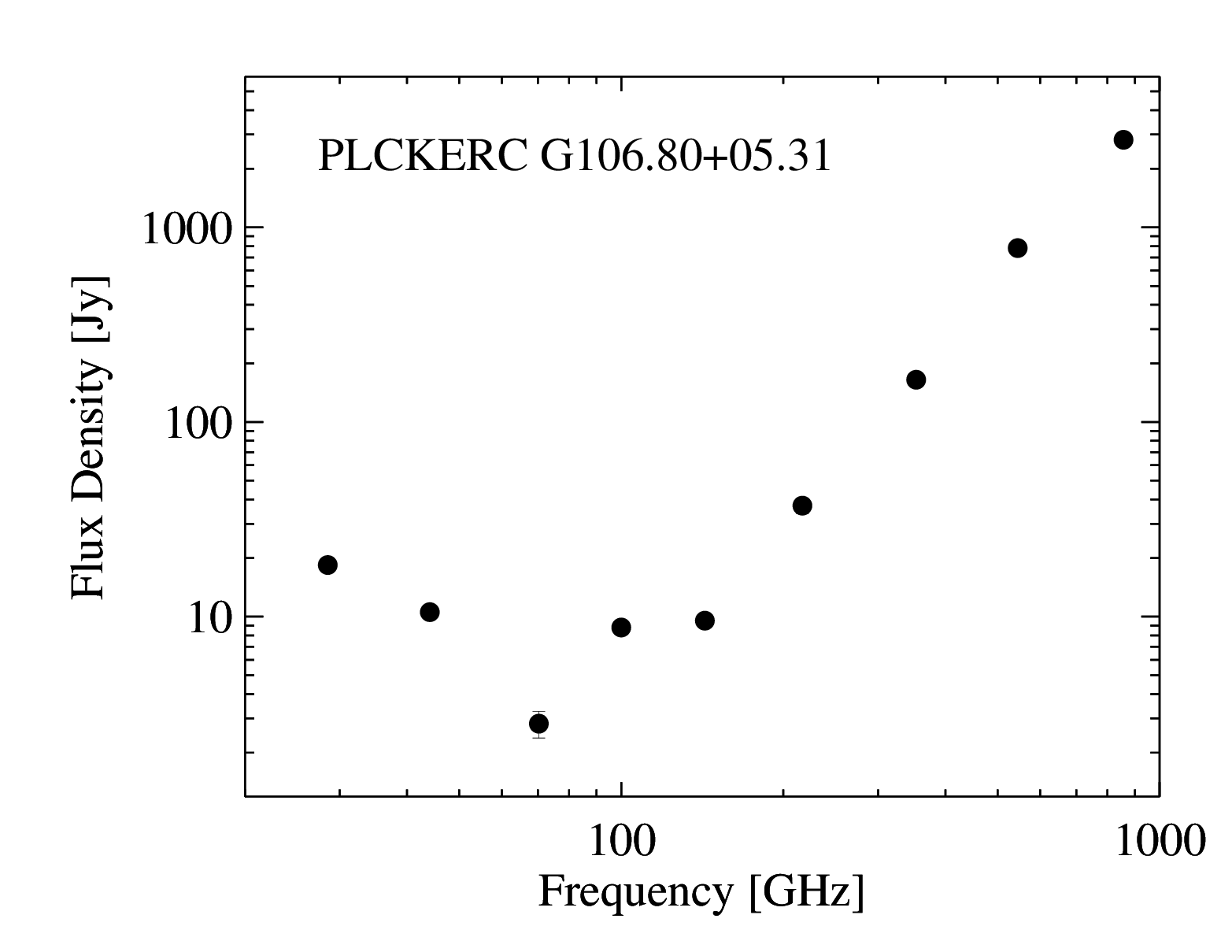} &  \hspace{-20pt}
\includegraphics[width=0.31\textwidth]{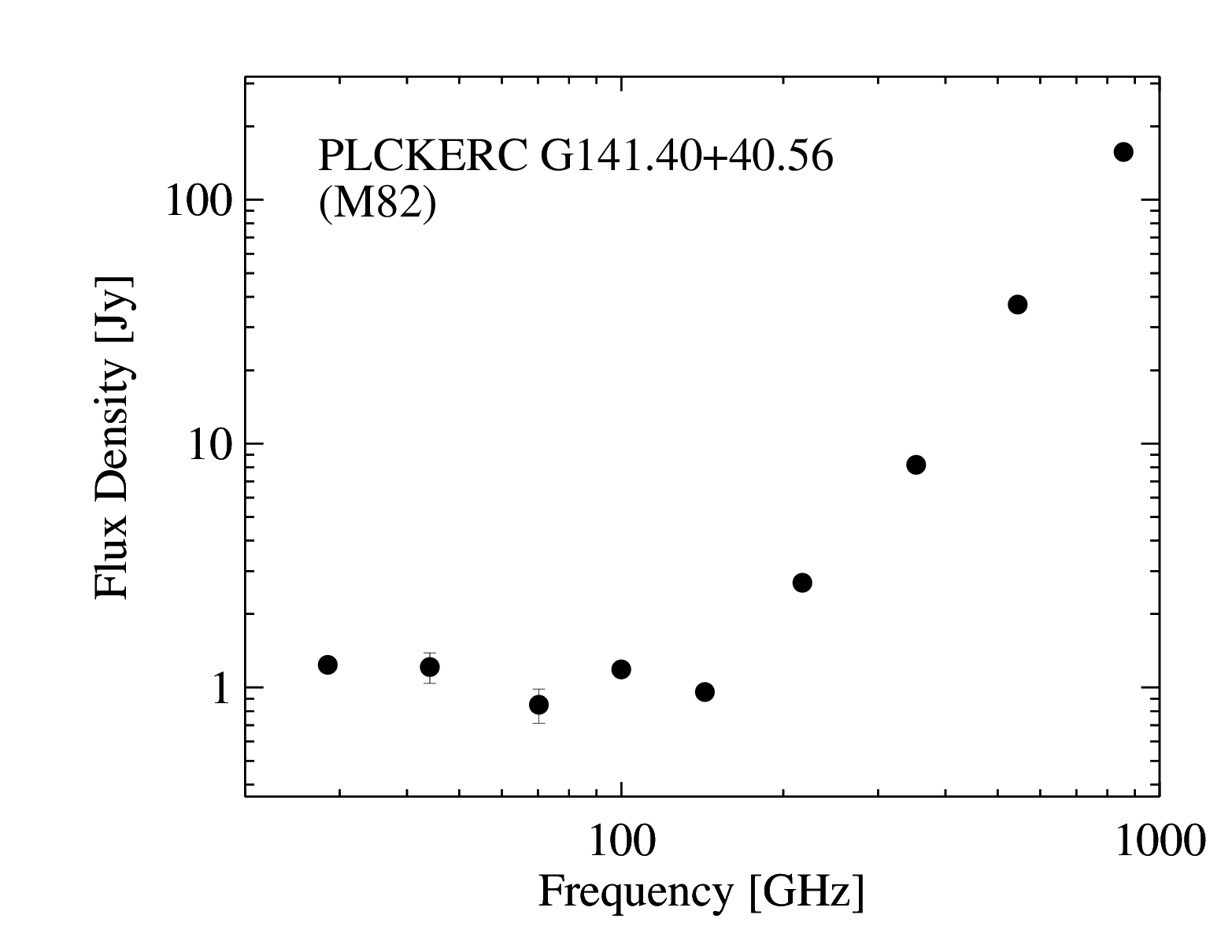} \\
\includegraphics[width=0.31\textwidth]{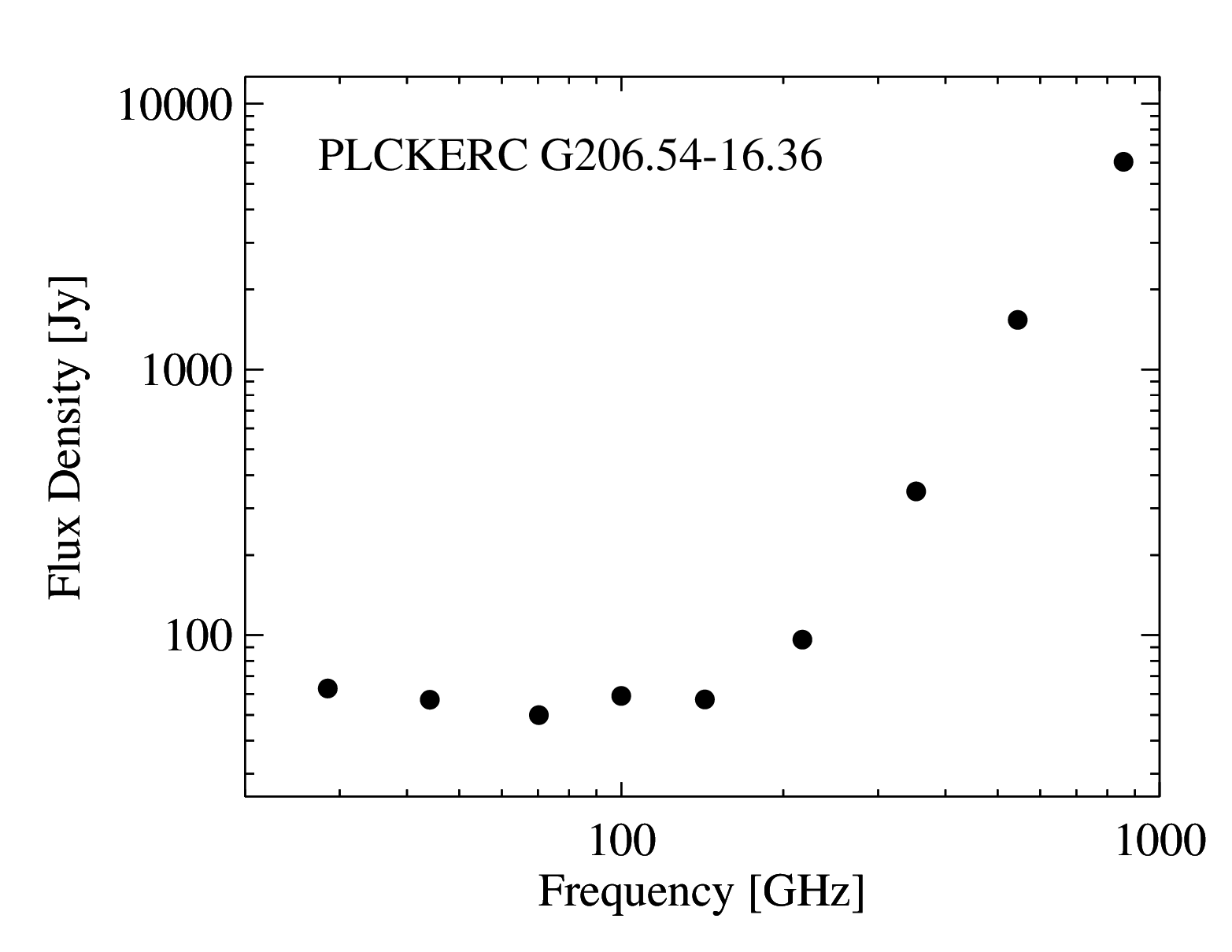} &  \hspace{-20pt}
\includegraphics[width=0.31\textwidth]{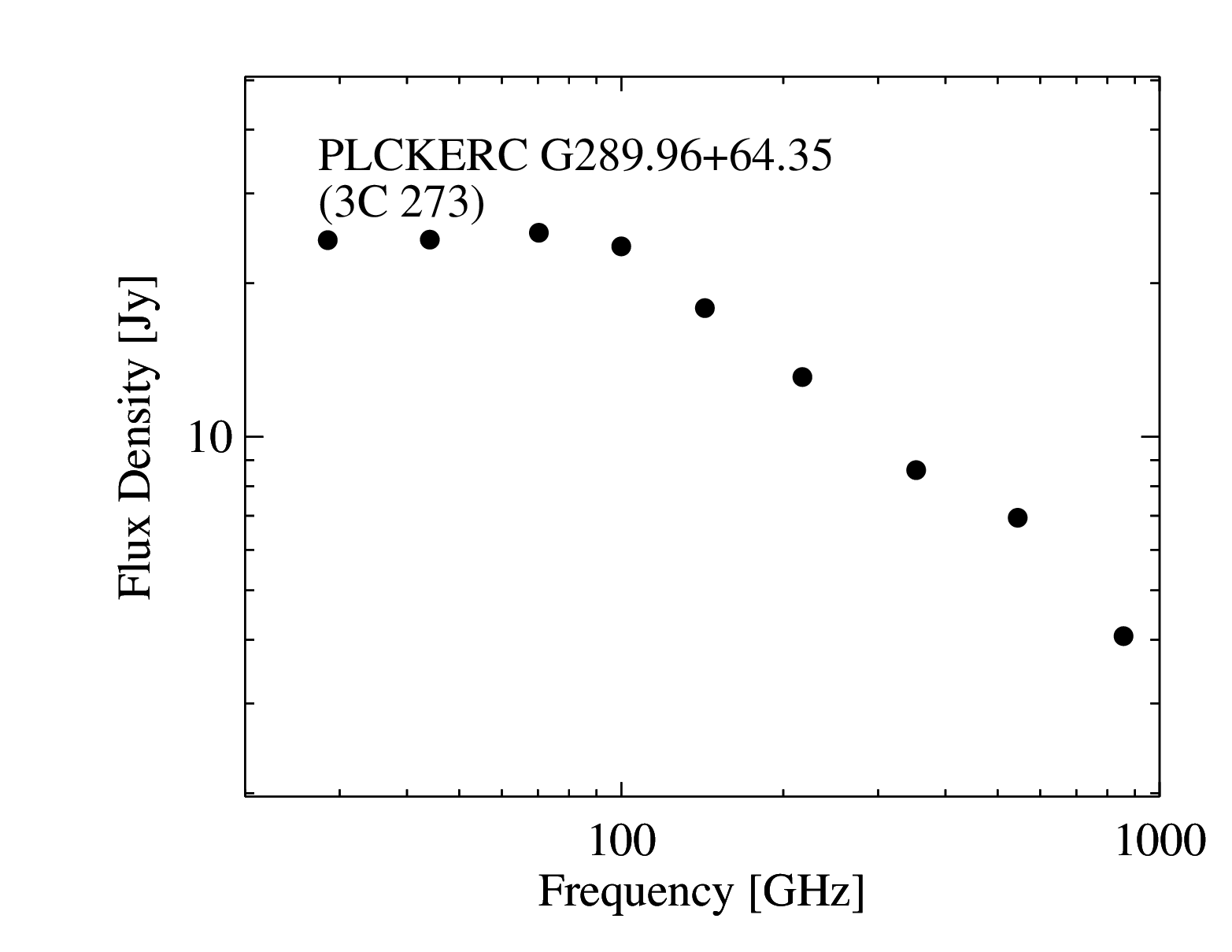} &  \hspace{-20pt}
\includegraphics[width=0.31\textwidth]{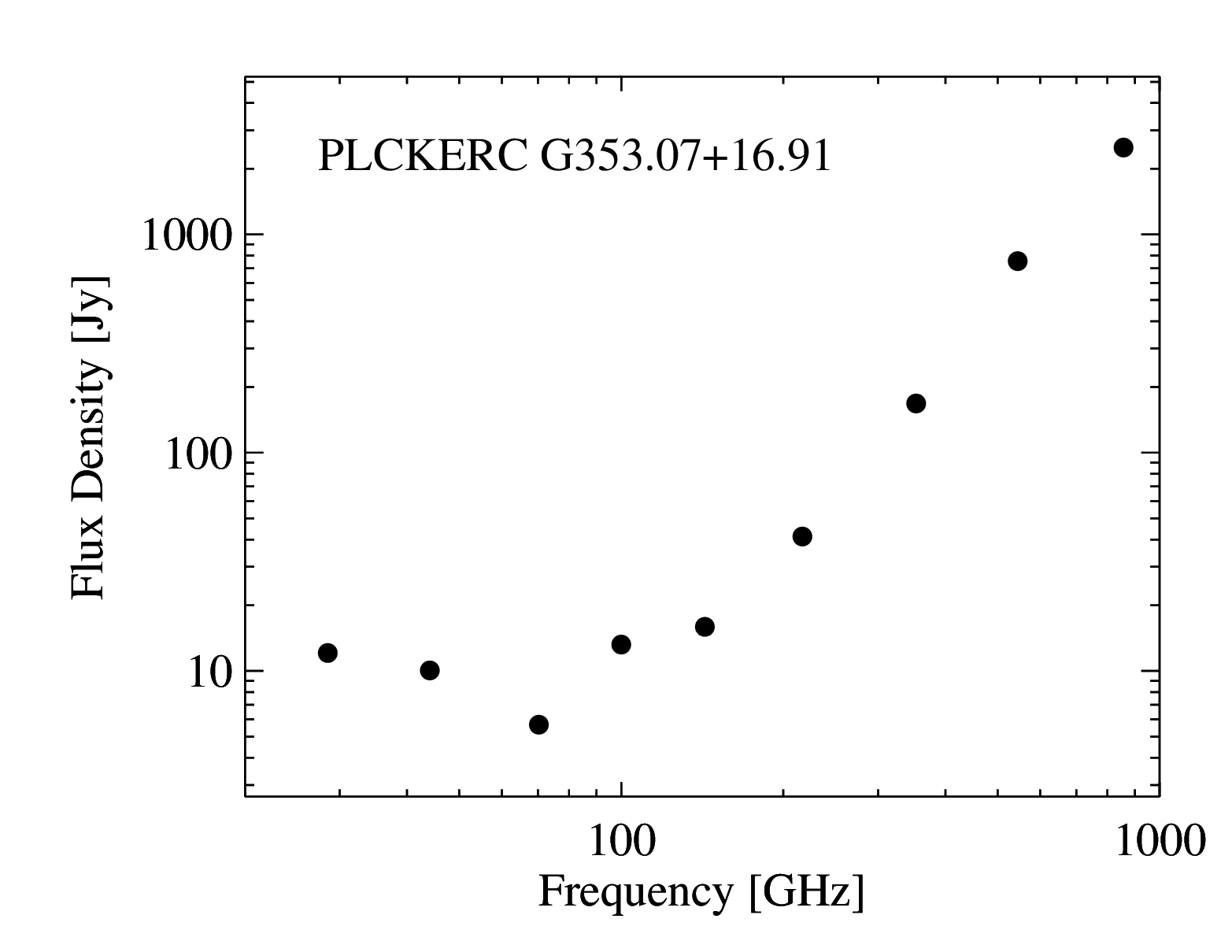} 
\end{tabular}
\caption{Sample SEDs of sources with detections in all nine \textit{Planck} frequency channels. Error bars are generally not visible because these bright sources have small flux-density uncertainties. A flux-density excess is evident at 100\,GHz for five of the six sources shown here (all except 3C\,273) which appears to be due to CO along the line of sight to the source.
\label{fig:sedall9}
}
\end{figure*}

\subsection{Detections in all nine bands} 
\label{sec:all9}
There are 79 sources that are detected at all nine frequency channels of \textit{Planck}. Most of these sources are Galactic, and only 12 are at $|b| > 30\degr$, including the well-known quasars 3C\,273 and 3C\,279, and the starburst galaxy M82. SEDs for some of these sources are shown in Figure~\ref{fig:sedall9}. Most of them show a visible bump in the SED at 100\,GHz. This is because the CO $J$=1$\rightarrow$0 rotational transition line at 115\,GHz falls within the bandpass of the \textit{Planck} 100\,GHz channel \citep{planck2011-1.7}, and thus the 100\,GHz flux densities of ERCSC sources are boosted by the CO emission. We use the bandmerged catalogue to estimate the flux-density excess due to CO contribution in Section~\ref{sec:co}, where we discuss the possibility of using this flux excess to trace the CO distribution at high Galactic latitudes.

\subsection{Detections in \textit{only} one band}
 
\begin{figure*}
\centering
 \begin{tabular}{ccc}
\includegraphics[width=0.31\textwidth]{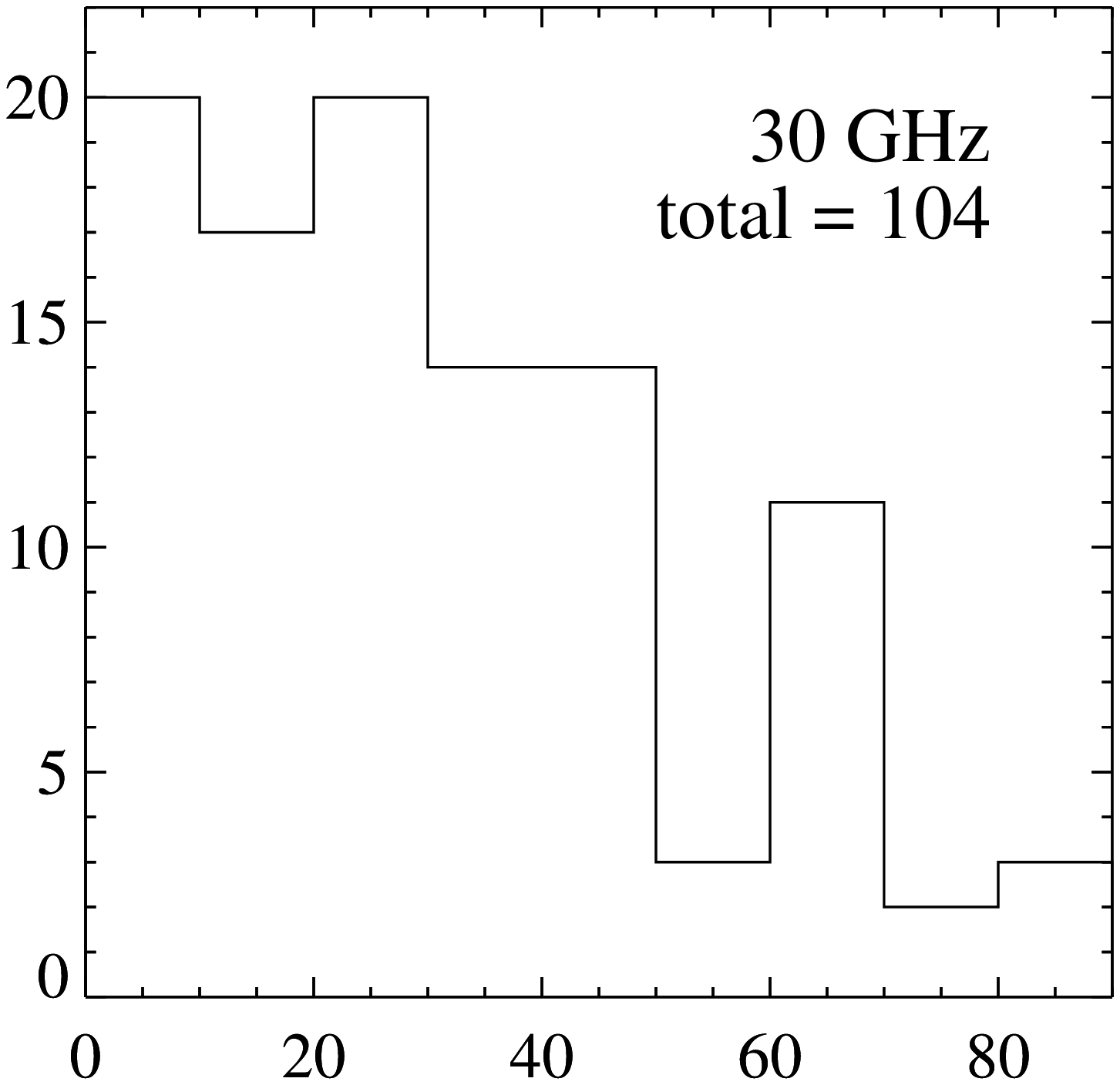} & \hspace{-30pt}
\includegraphics[width=0.31\textwidth]{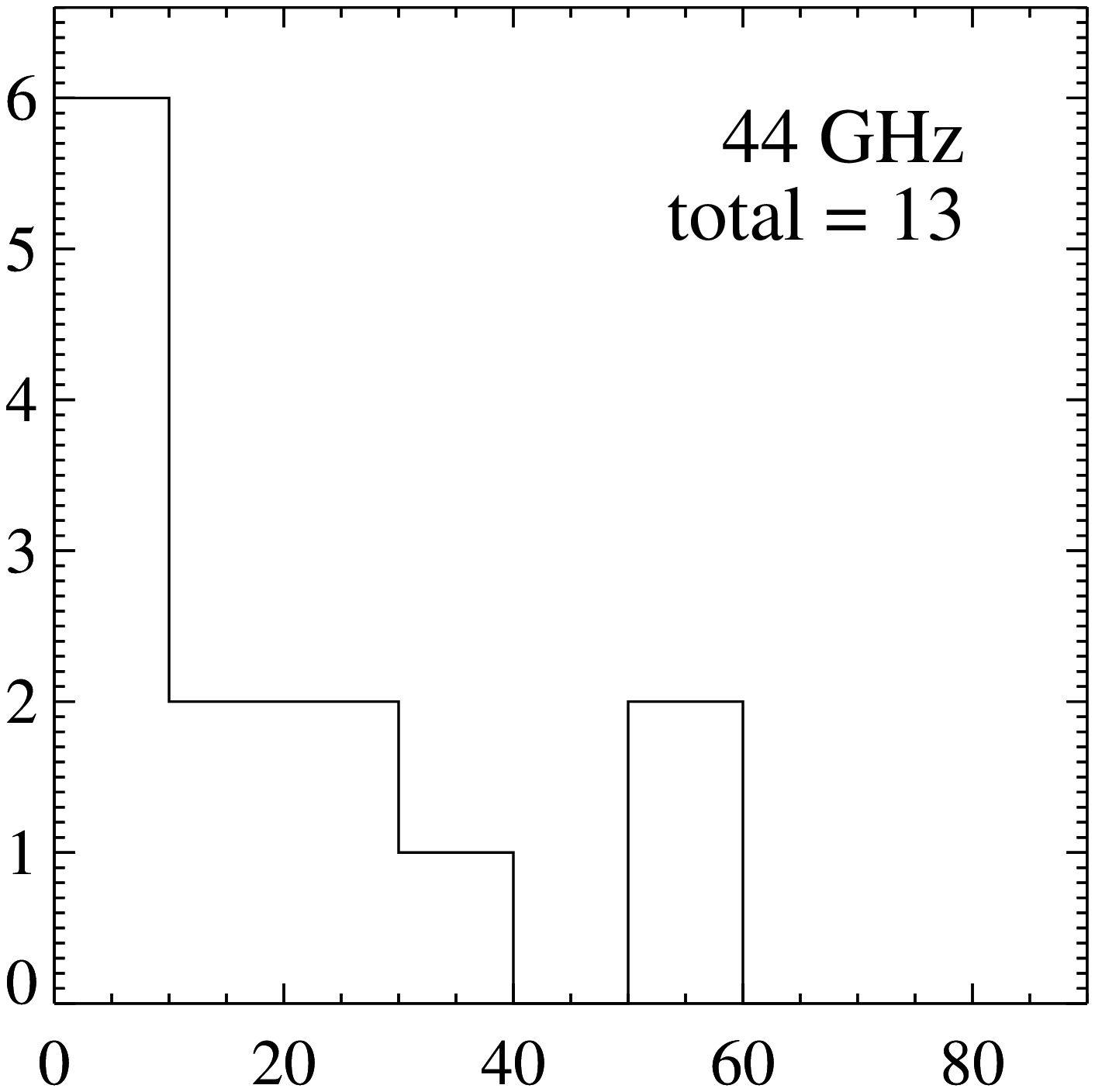} & \hspace{-30pt}
\includegraphics[width=0.31\textwidth]{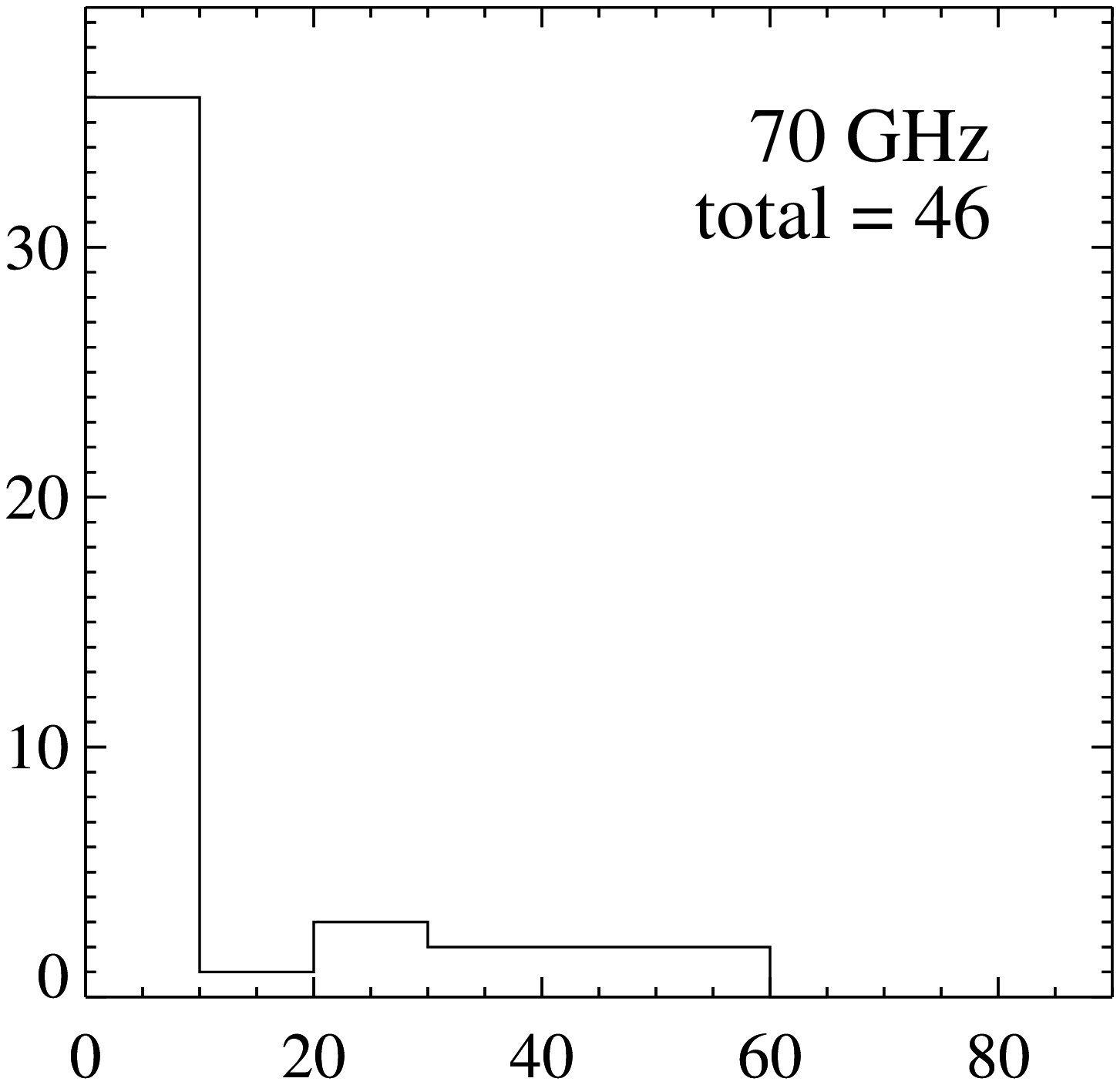} \\
\includegraphics[width=0.31\textwidth]{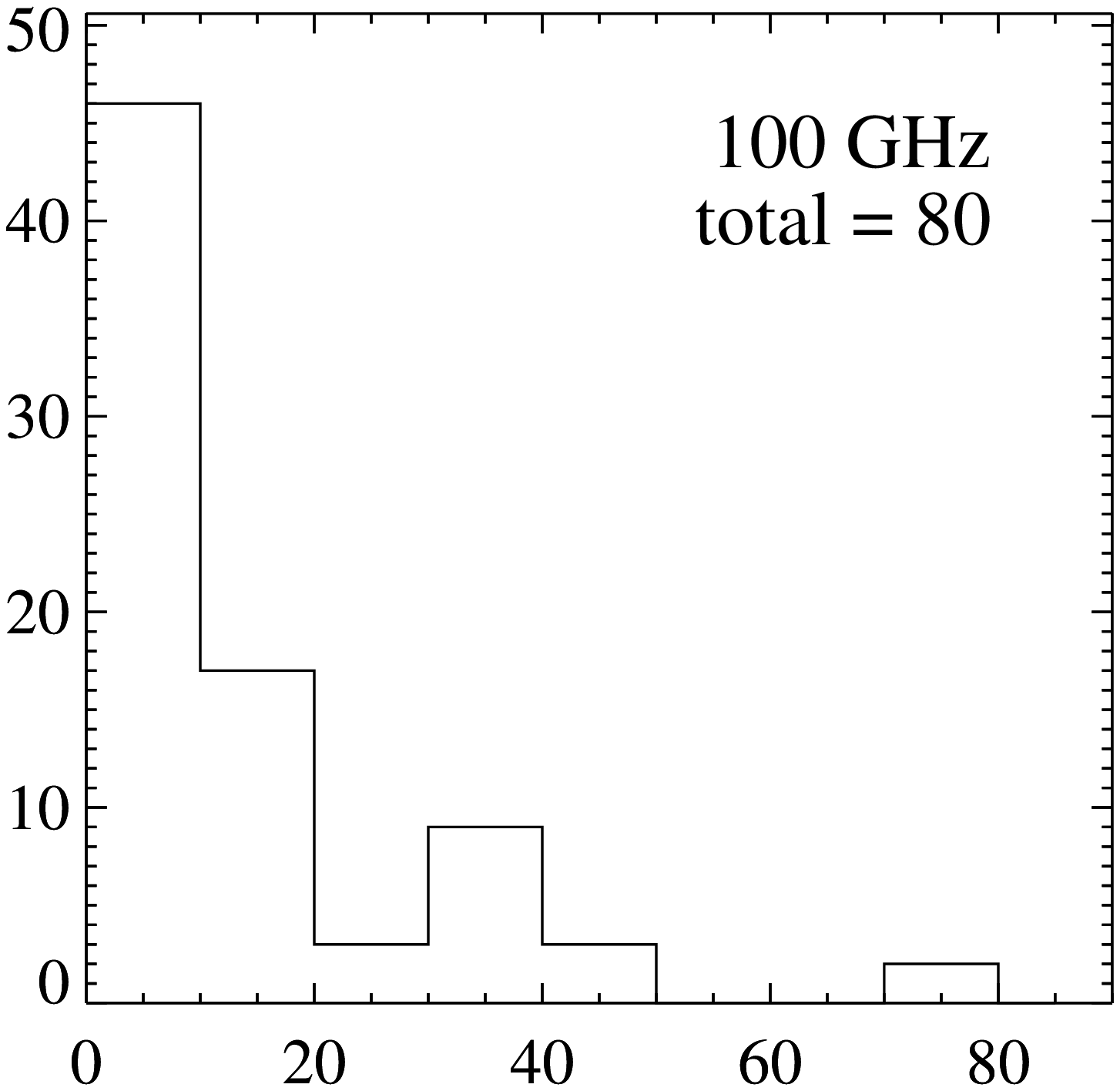} & \hspace{-30pt}
\includegraphics[width=0.31\textwidth]{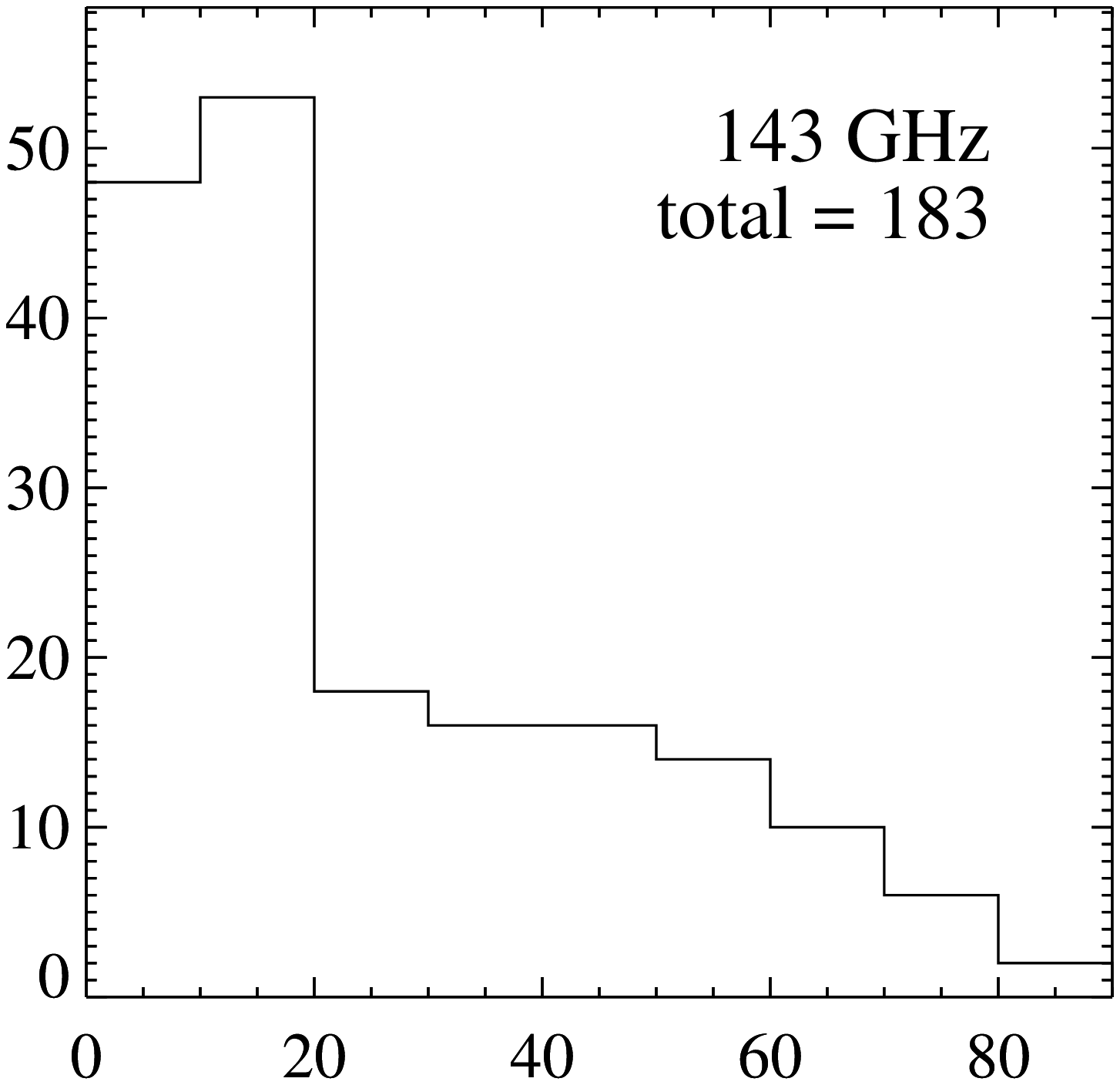} & \hspace{-30pt}
\includegraphics[width=0.31\textwidth]{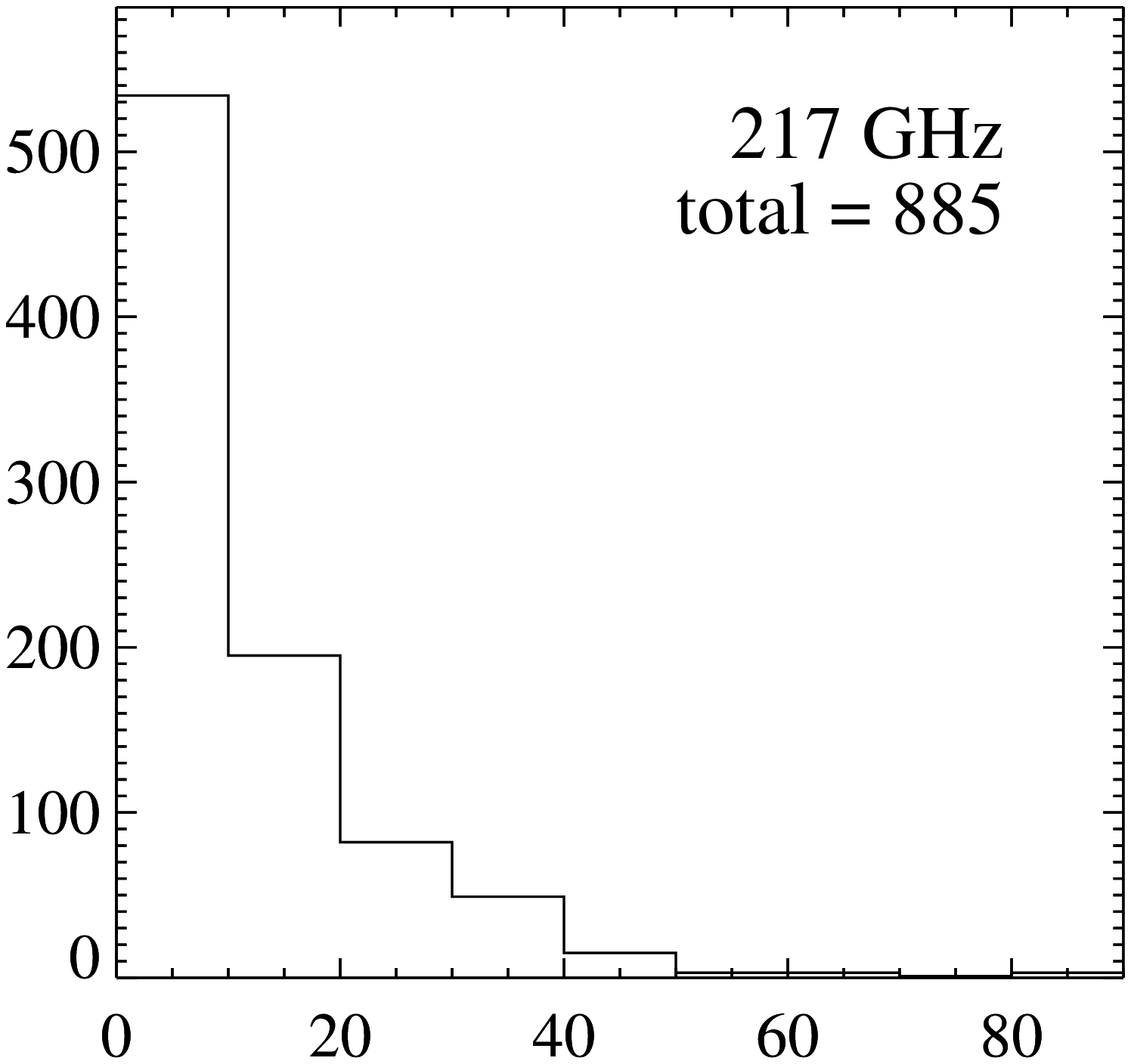} \\
\includegraphics[width=0.31\textwidth]{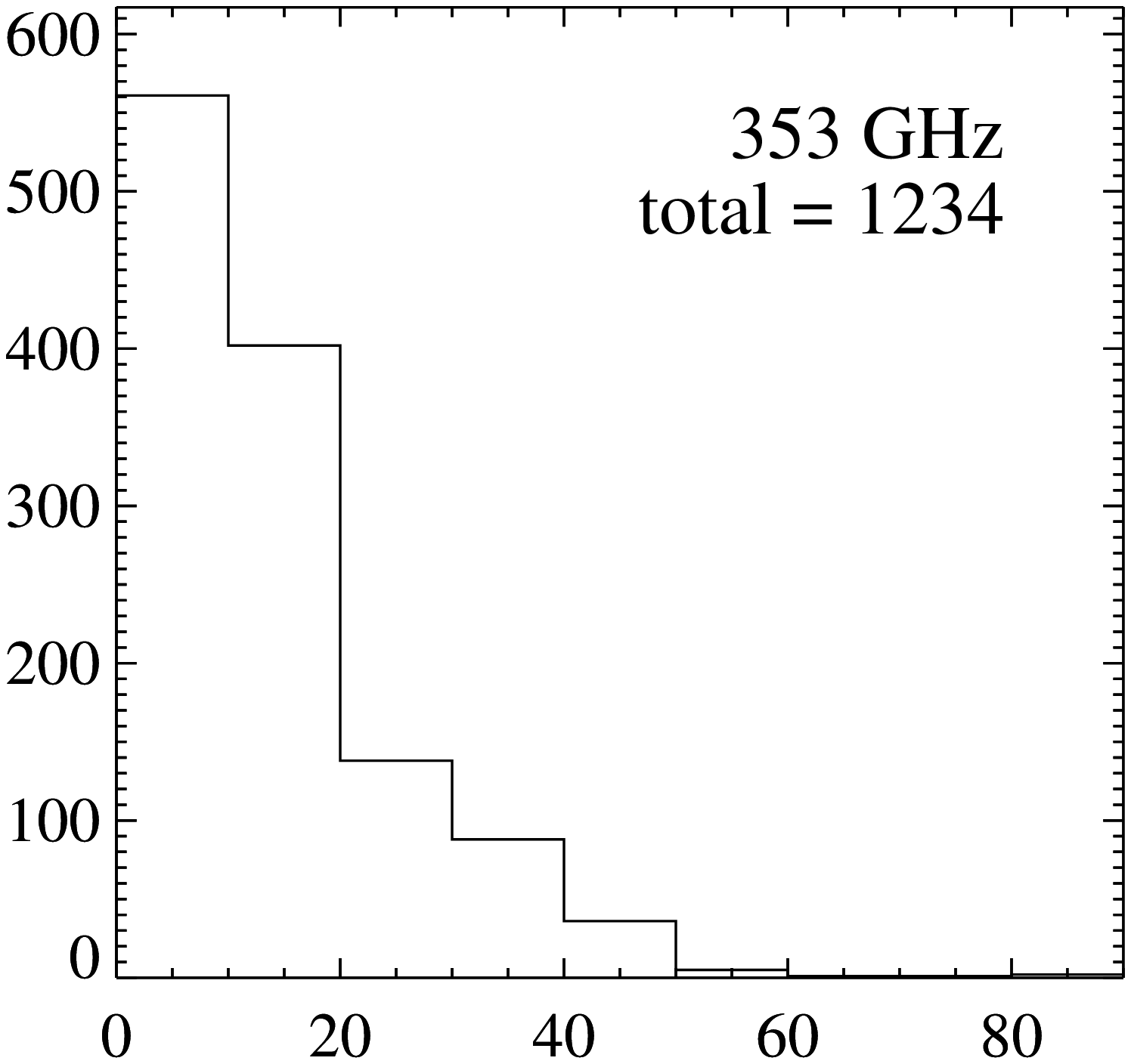} & \hspace{-30pt}
\includegraphics[width=0.31\textwidth]{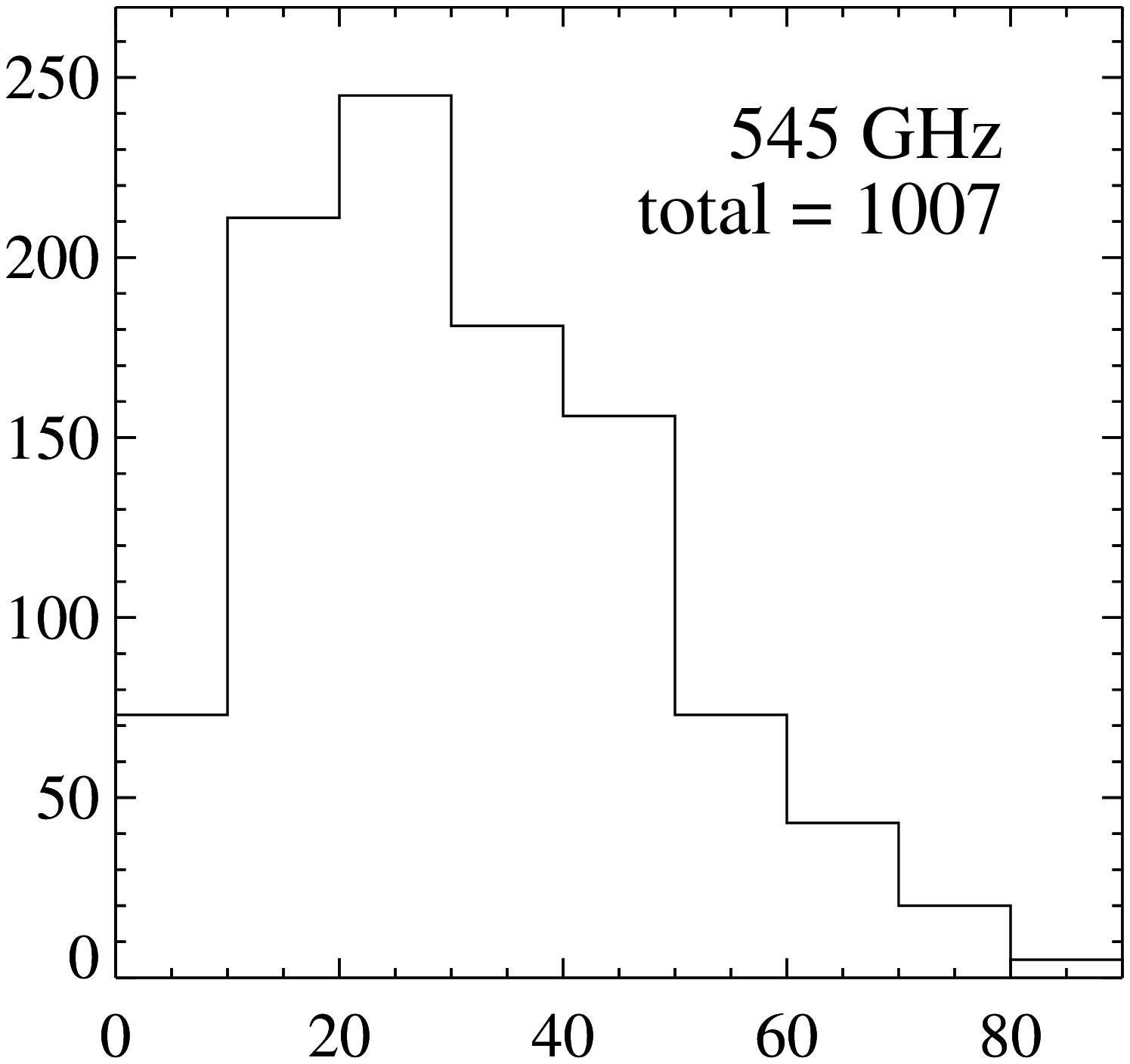} & \hspace{-30pt}
\includegraphics[width=0.31\textwidth]{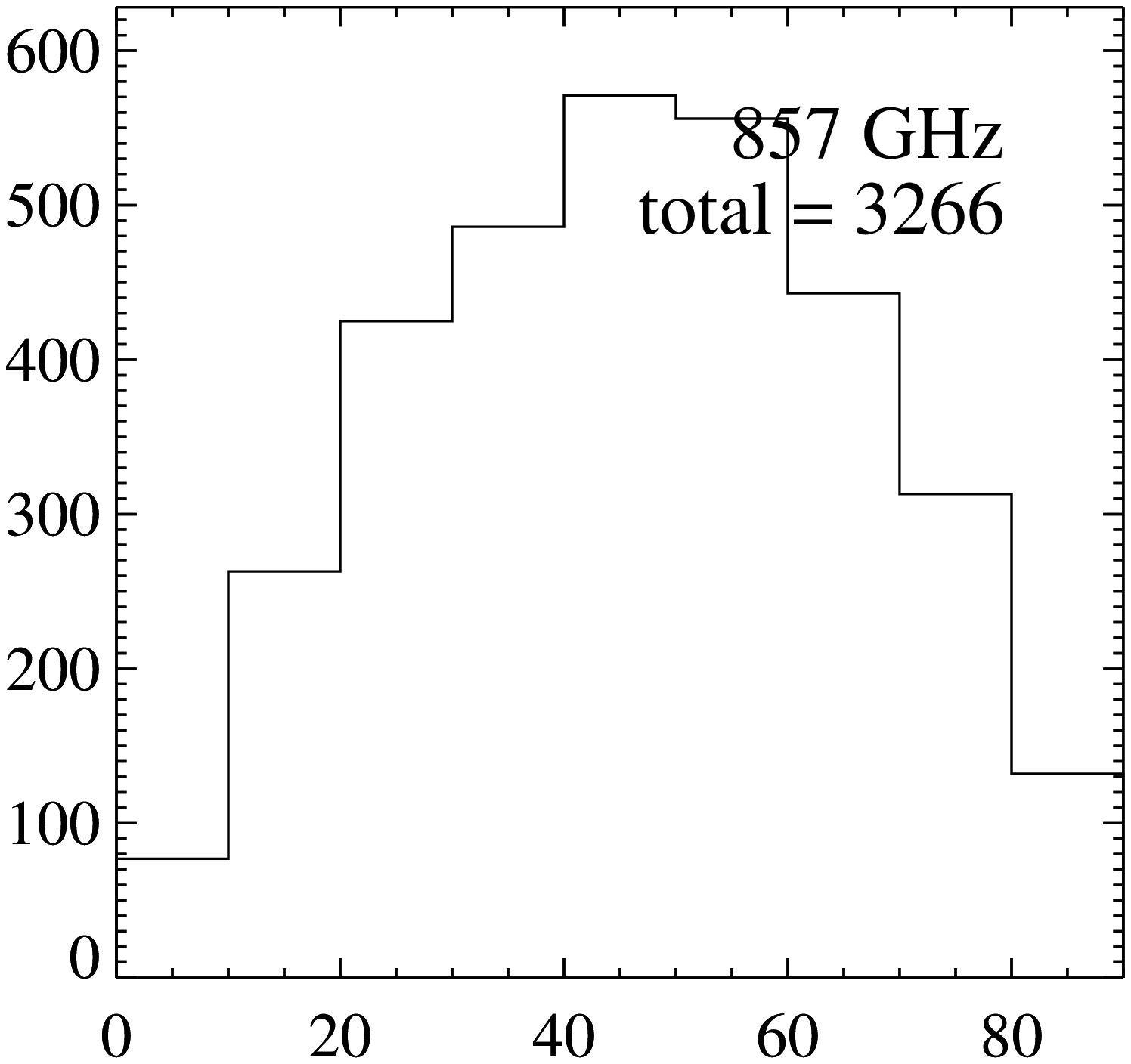} \\
\end{tabular}
\put(-235, -220){\large{$|$b$|$ [deg]}}
\put(-460, -60){\rotatebox{90}{\large{Number of sources}}}
\caption{Spatial distribution of sources as a function of Galactic latitude ($|b|$) that are detected in only one \textit{Planck} frequency channel. 
At lower frequencies, the sources are predominantly Galactic, while at higher frequencies the single-band sources extend to higher
latitudes which might suggest either high-latitude cirrus or extragalactic sources.  The total number of singular detections in each frequency is displayed in the top right corner of the panel.
\label{fig:single}
}
\end{figure*}

\begin{figure}
\centering
\includegraphics[width=0.45\textwidth]{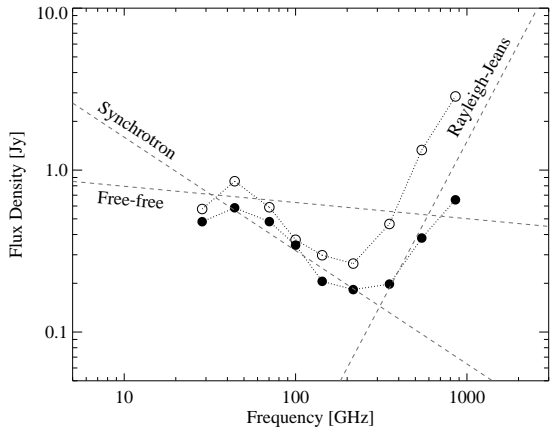} \\
\caption{The \textit{Planck} ERCSC flux density limit quantified as the faintest ERCSC source at $|b| < 10\degr$ (open circles) and at $|b| > 30\degr$ (filled circles), together with the expected spectra of known sources of foreground emission (dashed lines), including a $S_{\nu} \propto \nu^{-0.7}$ synchrotron component, a $\nu^{-0.1}$ free-free component, and a Rayleigh-Jeans component. \label{fig:sensitivity}}
\end{figure}

There are 6818 sources in the bandmerged catalogue that are detected in only one \textit{Planck} frequency channel.  Figure~\ref{fig:single} gives the distribution of these sources in Galactic latitude for each of the \textit{Planck} bands. The detection of a source in one band but not another can be due both to the spectral shape of the source and to the relative sensitivities across different \textit{Planck} bands \citep{planck2011-1.10}. In Table \ref{tab:ercsc}, we show the flux densities of the faintest sources detected at $|b| < 10 \degr$ and at $|b| > 30 \degr$, to quantify the sensitivity at each frequency. We also calculate the spectral indices between adjacent bands corresponding to these flux-density limits. Figure \ref{fig:sensitivity} plots the spectral shape derived from these flux-density limits, along with the expected spectra of the dominant radiation processes in the frequency regime covered by \textit{Planck}. It is clear that the 30\,GHz detections that are dominated by steep synchrotron spectra would be too faint to be detected at 44 or 70\,GHz, hence the large number of unmatched sources at 30\,GHz. The 44\,GHz band is the least sensitive of the \textit{Planck} bands, therefore if a source is detected in this band, it is likely to be detected at the 30\,GHz band regardless of spectral shape, hence the small number of unmatched sources. With the exception of the 30\,GHz channel, these singular sources are predominantly Galactic at frequencies up to 353\,GHz. This is because extragalactic sources in this frequency regime usually  have flat or close-to-flat spectra, as shown in Figure~\ref{fig:ccplot}, and are therefore more likely to be detected at neighbouring frequencies. At higher frequencies, both the Galactic and the extragalactic source populations are dominated by dusty sources with rising spectra. Since the detection limit in these bands also mimics a rising spectrum (Fig. \ref{fig:sensitivity}), the relative steepness of the two slopes determines whether a source is detected in one band or another. For example, the detection limit rises as $\nu^{1.7}$ for sources at $|b| < 10 \degr$ between 217 and 857\,GHz. If a source has $\alpha > 1.7$, it is likely to be detected in the 857\,GHz band but missed in the 217\,GHz band.

Although the balance of map sensitivity and source spectral shape is the major reason for these single detections, it is not the only reason:  the ERCSC is not a complete catalogue (to ensure a 90$\%$ cumulative reliability at each band, some detections were removed from the final source list due to the noisy sky background even though they individually passed the detection threshold; see \citealt{planck2011-1.10}); and it is also possible that  a few of the detections are spurious.
 
\begin{table*}
\begin{minipage}{140mm}
\caption{\textit{Planck} ERCSC flux density limit\label{tab:ercsc}}
\begin{tabular}{lccccccccccc}
\hline
Frequency [GHz] & 30 & 44 & 70 & 100 & 143 & 217 & 353 & 545 & 857 \\
\hline
\omit&\multicolumn{9}{c}{$|b| < 10\degr$}\\
Flux density limit\tablefootmark{a} [Jy] & 0.575 & 0.853 & 0.589 & 0.371 & 0.298 & 0.265 & 0.465 & 1.331 & 2.850 \\
Spectral index \tablefootmark{b} &  & 0.90 & $-$0.79 & $-$1.31 & $-$0.61 & $-$0.28 & 1.16 & 2.42 & 1.68 \\ 
\hline
\omit&\multicolumn{9}{c}{$|b| > 30\degr$}\\
Flux density limit\tablefootmark{a} [Jy] & 0.480 & 0.585 & 0.481 & 0.344 & 0.206 & 0.183 & 0.198 & 0.381 & 0.655 \\
Spectral index \tablefootmark{b} &  & 0.45 & $-$0.42 & $-$0.95 & $-$1.44 & $-$0.28 & 0.16 & 1.51 & 1.20 \\
\hline
\end{tabular}\\
\medskip
\tablefoottext{a}{Flux density of the faintest source at $|b| < 10\degr$ or $|b| > 30\degr$ in the ERCSC. }\\
\tablefoottext{b}{Spectral index $\alpha$, calculated based on the flux density limit of the adjacent bands, $S_{\nu} \propto \nu^{\alpha}$. For the LFI bands, the central frequencies of 28.5, 44.1 and 70.3\,GHz are used in the calculation instead of the nominal frequencies.}
\end{minipage}
\end{table*}

\section{Probing CO at High Galactic Latitudes?}
\label{sec:co}

As we mentioned in Section~\ref{sec:all9}, the \textit{Planck} 100\,GHz flux density measurements can be contaminated by the $^{12}$CO $J$=1$\rightarrow$0 line at 115.271\,GHz \citep{planck2013-p03a}. This CO is probably Galactic,
along the line of sight between the \textit{Planck} satellite and the source. However, an origin within the source cannot be 
ruled out since even  a modest redshift  ($z\sim0.1$) would shift the line further inside the 100\,GHz bandpass.
We take two approaches in assessing the flux-density excess due to the CO contamination. 

In the first approach (Approach I), we select bandmerged sources that were detected at 100\,GHz, at two or more channels below 100\,GHz, and at two or more channels above 100\,GHz. This allows us to fit three types of simple models to the source SED: 

\noindent i) single power law 
\begin{equation}
\log S = p_0 + p_1 \log\nu, 
\label{eqn:sp}
\end{equation}
ii) quadratic fit
\begin{equation}
\log S = p_0 + p_1 \log\nu + p_2 (\log\nu)^2, 
\end{equation}
iii) double power law%\footnote{Require detections in at least 5 Planck channels excluding the 100\,GHz one.}
\begin{equation}
S = p_0 (\nu/\nu_0)^{p_1} + p_2 (\nu/\nu_0)^{p_3},
\end{equation}
where $p_i (i = 0, 1, 2, 3)$ are parameters that are different in each model.
The 100\,GHz flux densities are excluded in the SED fitting. A model fit is accepted when the $\chi^2$ values indicate that the model is compatible with $95\%$ of the data. In cases when more than one model was acceptable, we selected the model with fewest parameters unless an additional parameter significantly improved the value of the reduced $\chi^2$ (as described in
\citealt{Bevington2003}).

\begin{figure*}
\centering
\includegraphics[width=0.45\textwidth]{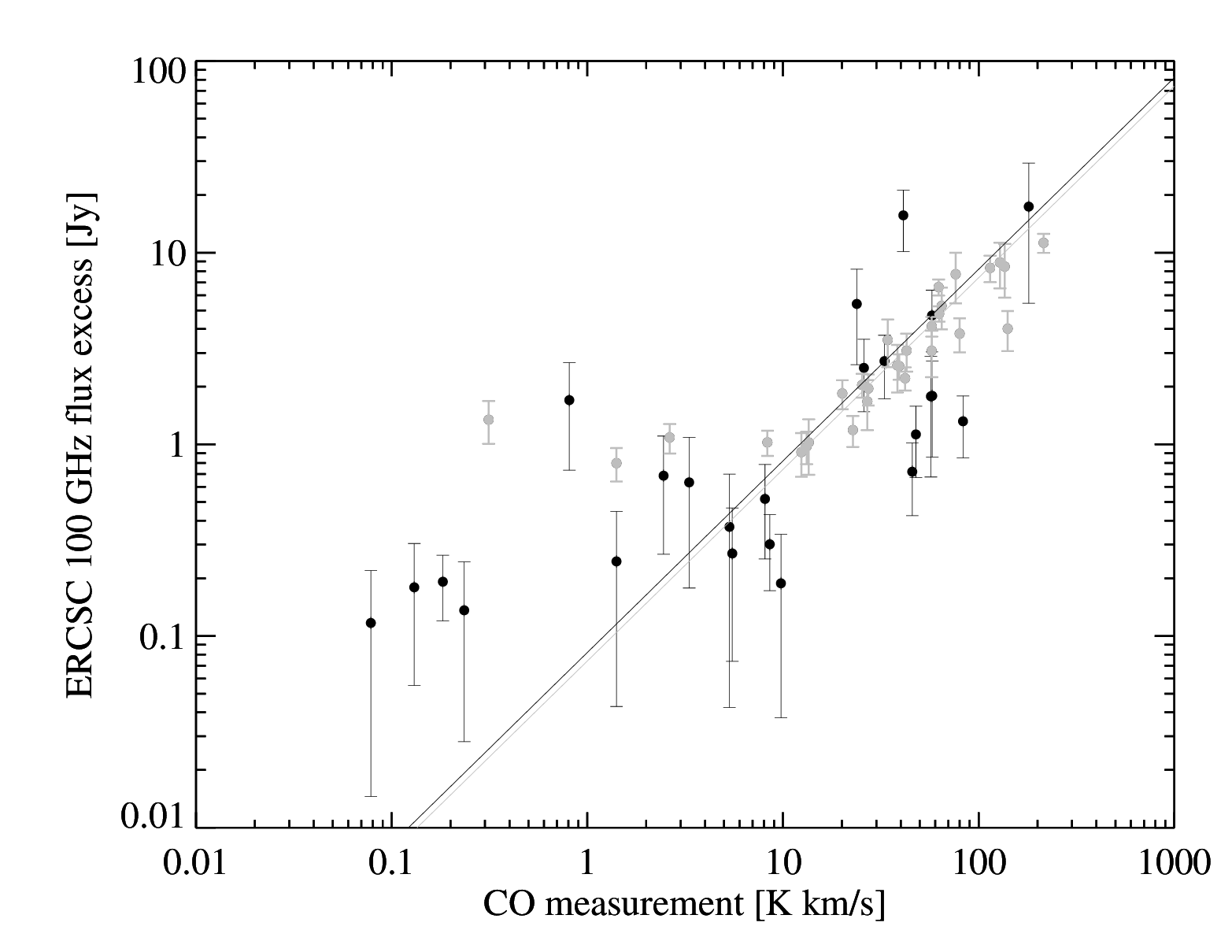}
\includegraphics[width=0.35\textwidth,angle=90]{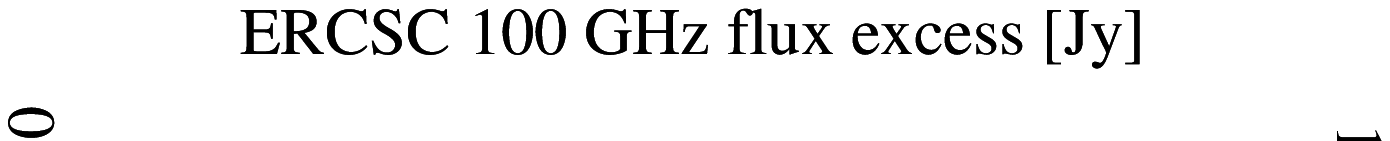} 
\caption{Comparison of ERCSC 100 flux-density excess determined by Approach I (left) and Approach II (right) with the CO intensity from the \citet{dame2001} Galactic CO map. The filled circles are all the sources with $S_{\rm excess} > \sigma_{\rm excess}$, while red (or light colored circles) indicates sources with $S_{\rm excess} > 3\sigma_{\rm excess}$. The black line is a fit to all the sources while the red line is fitted only to sources with highly significant excess. The agreement in the fits between these two approaches indicates that we are obtaining a robust measure of the CO emission along the line of sight to these sources. \label{fig:co}}
\end{figure*}

There are 367 sources in the bandmerged catalogue that meet the selection criteria, out of which 225 (61.3$\%$) yield a good fit.  For the sources that have a good fit, we subtract the fitted values at 100\,GHz from the ERCSC 100\,GHz flux densities, i.e., $S_{\rm excess} = S_{100}^{\rm ERCSC} - S_{100}^{\rm model}$, and the uncertainty of this excess is $\sigma_{\rm excess} = \sqrt {(\sigma_{100}^{\rm model})^2 + (\sigma_{100}^{\rm ERCSC})^2}$. We consider a source to have a positive flux-density excess at 100\,GHz when $S_{\rm excess} > \sigma_{\rm excess}$; 123 sources are found to have a positive 100 GHz flux density excess. Assuming this excess comes entirely from the CO $J$=1$\rightarrow$0 rotational transition line, we would expect a linear correlation between the amplitude of the excess and the observed CO intensity from \citet{dame2001}.  In Figure \ref{fig:co} (left panel), we plot the excess values against the CO intensity  for the 51 ERCSC sources that are in the region covered by the \citet{dame2001} CO map. There is good agreement, especially in the high CO intensity zone. We then apply a linear fit in log-log space, forcing the slope to be 1, i.e., $\log S_{100}^{\rm excess} = \log I_{\rm CO} + b$. For all the points with $S_{\rm excess} > \sigma_{\rm excess}$, the intercept $b$ has a value of $-1.086\pm0.013\pm0.452$. If we restrict the fit to only the points with highly significant excess (i.e., $S_{\rm excess} > 3\sigma_{\rm excess}$), we find an intercept value of $-1.131\pm0.014\pm0.105$. Here, the first error term is propagated from $\sigma_{\rm excess}$, the second error term is the scattering of $\log S_{100}^{\rm excess} - \log I_{\rm CO}$ with respect to the mean value of b. 

\begin{figure*}
\centering
\includegraphics[width=0.45\textwidth]{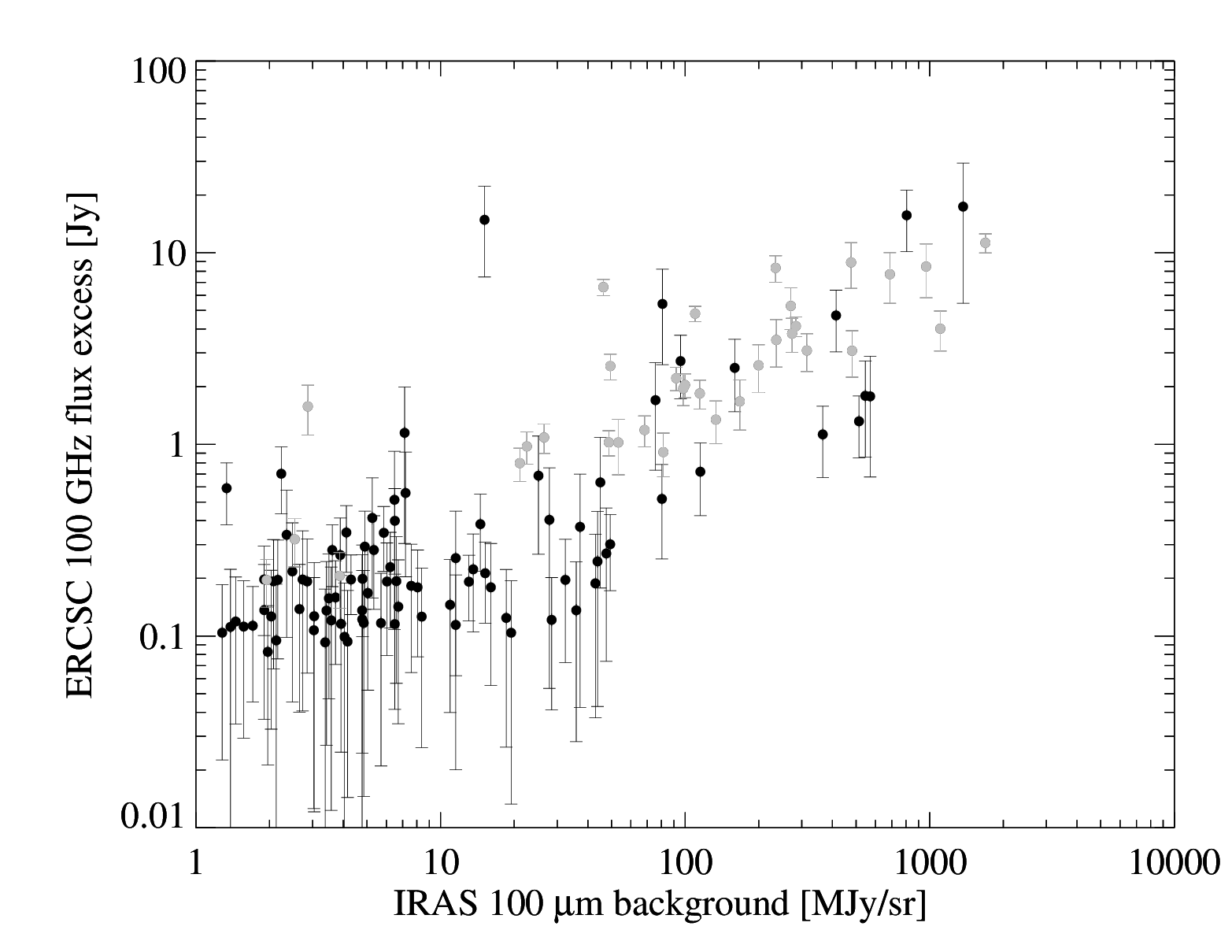} 
\includegraphics[width=0.35\textwidth,angle=90]{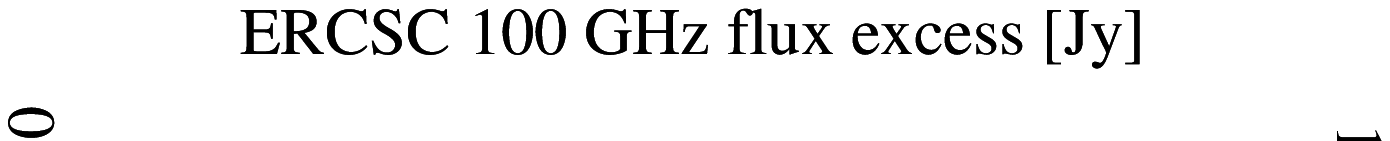} 
\caption{Comparison of ERCSC 100\,GHz flux excess determined by Approach I (left) and Approach II (right) with the intensity from the IRAS 100$\mu$m map. Filled circles show the sources with $S_{\rm excess} > \sigma_{\rm excess}$, while red (or light colored circles) indicates sources with $S_{\rm excess} > 3\sigma_{\rm excess}$.  \label{fig:iras}}
\end{figure*}

In a second approach (Approach II), we select sources that have detections at 70, 100 and 143\,GHz. Assuming a single power-law spectrum across these three bands, we interpolate between the 70 and 143\,GHz flux densities to approximate the true 100\,GHz flux density of the source. There are 482 sources in the bandmerged catalogue that meet this selection criterion, out of which 147 are at $|b|>30\degr$. We apply the same definition for sources with flux excess or significant flux excess in the 100\,GHz channel and find 254 sources have positive flux excess at 100\,GHz, out of which 53 are at $|b|>30\degr$. We again compare the excess values of the ERCSC sources with ground-based CO measurements. Of the 254 sources, 154 reside in the region covered by the \citet{dame2001} CO map. We see in Figure~\ref{fig:co} (right panel) the linear correlation between this set of flux-density excess values and the measured CO intensity. In this case we find an intercept of $-1.206\pm0.008\pm0.355$ for all the sources with positive excess and $-1.198\pm0.008\pm0.225$ for sources with highly significant excess. 

\begin{table*}
\begin{minipage}{140mm}
\caption{List of sources with 100\,GHz flux excess in the \textit{Planck} ERCSC (Only the first 10 rows are shown here. The full table is available online.)
\label{tab:sample}}
\begin{tabular}{lcccccc}
\hline
Name & RA (J2000)    & Dec (J2000)    & $S_{70\,GHz}$ & $S_{100\,GHz}$ & $S_{143\,GHz}$ & $S_{excess}$  \\
  & [h m s] & [d m] & [Jy]         & [Jy]          & [Jy]          & [Jy]          \\
\hline
 PLCKERC G000.65-00.03     & 17 47 17    & -28 23   &  109.28$\pm$ 9.02  &  227.76$\pm$ 9.90  &  196.90$\pm$ 7.13  &   81.39$\pm$11.91  \\
 PLCKERC G002.43+05.84     & 17 29 21    & -23 45   &    1.33$\pm$ 0.14  &    1.77$\pm$ 0.08  &    1.57$\pm$ 0.06  &    0.32$\pm$ 0.12  \\
 PLCKERC G005.89-00.40     & 18 00 34    & -24 03   &   27.05$\pm$ 1.17  &   30.48$\pm$ 0.65  &   30.64$\pm$ 0.80  &    1.71$\pm$ 0.98  \\
 PLCKERC G005.97-01.14     & 18 03 34    & -24 21   &   72.59$\pm$ 2.83  &   60.23$\pm$ 1.47  &   37.16$\pm$ 1.76  &    8.16$\pm$ 2.17  \\
 PLCKERC G008.13+00.22     & 18 02 59    & -21 48   &    7.55$\pm$ 1.20  &   11.15$\pm$ 1.26  &   10.08$\pm$ 0.89  &    2.43$\pm$ 1.49  \\
 PLCKERC G009.34-19.61     & 19 24 52    & -29 14   &   12.49$\pm$ 0.19  &   11.08$\pm$ 0.08  &    9.13$\pm$ 0.07  &    0.39$\pm$ 0.12  \\
 PLCKERC G010.18-00.35     & 18 09 27    & -20 18   &   46.24$\pm$ 2.34  &   50.14$\pm$ 2.32  &   45.38$\pm$ 2.36  &    4.33$\pm$ 2.86  \\
 PLCKERC G012.17-05.71     & 18 33 41    & -21 03   &    2.49$\pm$ 0.18  &    2.09$\pm$ 0.07  &    1.54$\pm$ 0.04  &    0.13$\pm$ 0.11  \\
 PLCKERC G012.80-00.20     & 18 14 15    & -17 55   &   51.75$\pm$ 1.63  &   60.71$\pm$ 1.56  &   54.54$\pm$ 1.87  &    7.59$\pm$ 1.99  \\
 PLCKERC G014.22+42.21     & 15 50 38    & +05 27   &    1.50$\pm$ 0.18  &    1.48$\pm$ 0.07  &    1.06$\pm$ 0.04  &    0.22$\pm$ 0.10  \\
\hline
\end{tabular}

\medskip

\end{minipage}
\end{table*}

\begin{figure}
\centering
\includegraphics[width=0.49\textwidth]{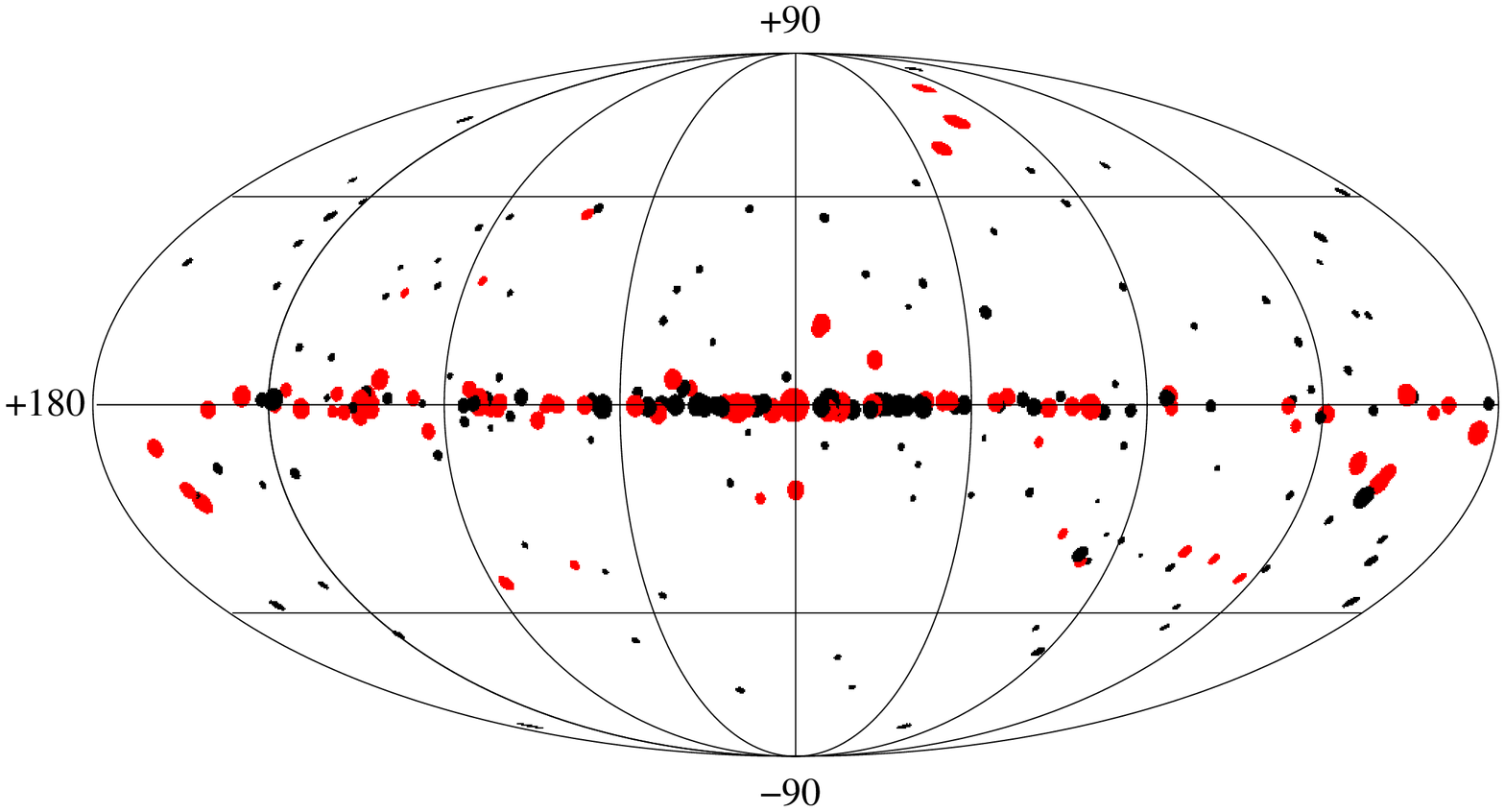} 
\caption{Sky distribution of ERCSC sources that have flux-density excess at 100\,GHz, as determined by Approach~II. The black filled circles are the sources with $S_{\rm excess} > \sigma_{\rm excess}$, while red indicates sources with $S_{\rm excess} > 3\sigma_{\rm excess}$. Sizes of the circles are scaled with the excess values. \label{fig:skymap}}
\end{figure}

The \citet{dame2001} CO map is largely restricted to the Galactic plane. Since the gas and dust content in the ISM are well correlated, we further compare the ERCSC 100\,GHz flux-density excess values with the median intensity within an aperture radius of 1~FWHM at the source position in the IRAS 100\,$\mu$m full-sky map\footnote{\href{http://irsa.ipac.caltech.edu/data/Planck/release\_1/external-data/external\_maps.html}{http://irsa.ipac.caltech.edu/data/Planck/release\_1/external-data/external\_maps.html} } . We see that the linear relationship extends over the entire sky (Fig.~\ref{fig:iras}) for excess values estimated by both approaches. This suggests that we can use the ERCSC 100\,GHz flux-density excess to probe the distribution of CO at high Galactic latitudes. Since the values fitted in the two approaches are not distinguishable given the large scatter, and there is little difference between the fits from all the sources and the sources with highly significant excess, we therefore decide to adopt Approach II and a linear relation of $\log S_{100}^{\rm excess} = \log I_{\rm CO} -1.206(\pm0.008\pm0.355)$ as it is a simpler approach and has a bigger sample size. The full sample of sources with a positive 100\,GHz flux density excess is given in Table \ref{tab:sample}, which list the positions of the sources, their 70, 100 and 143\,GHz flux densities in the ERCSC, and the excess flux densities at 100\,GHz. Figure \ref{fig:skymap} presents the sky distribution of this sample. All the sources with $S_{\rm excess} > \sigma_{\rm excess}$ are indicated by filled circles, while red highlights the sources with $S_{\rm excess} > 3\sigma_{\rm excess}$.

The most natural explanation would be that the excess CO emission is coming from the source itself. However, the vast majority of these sources are radio-loud extragalactic AGN,
especially at high latitude. The median redshift of the sources determined by cross-correlating with the NASA Extragalactic Database (NED) is $z\sim0.5$. At these redshifts,
the rest-frame 115 GHz CO line would be redshifted out of the 100 GHz \textit{Planck} bandpass. Only at $z<0.35$ would the CO line remain in the 100 GHz bandpass.
Furthermore, for even a $z\sim0.1$ source, a typical molecular gas mass of
10$^{10}$\,M$_{\sun}$ \citep{Frayer2011, Riechers2011},would result in an bandpass integrated flux density excess
at 100 GHz of less than a $\umu$Jy; this is many orders of magnitude below the typical photometric noise of the sources. Thus, it appears that the CO that we detect
is Galactic CO emission along the line of sight to the source.

It is important to note that the excess CO emission is measured in the photometry of the source which in turn is measured by subtracting the median sky background in an annulus
around the source. It therefore appears that the detected CO emission is in excess of the diffuse, spatially extended, CO emission and is clumpy on the scale of the \textit{Planck} 100 GHz
source size which corresponds to $\sim$10$\arcmin$ radius. This is a scale which has not been probed by previous ground-based, CO mapping experiments and this catalogue
provides a unique window into this clumpy emission. Furthermore, even the \textit{Planck} component separated CO maps are unable to trace this emission at the location
of the sources since the maps suffer from leakage of source flux into the component maps. 

\begin{figure}
\centering
\includegraphics[width=0.35\textwidth,angle=90]{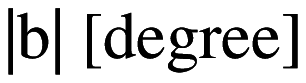} 
\caption{The distribution of ERCSC 100\,GHz flux excess with respect to the galactic latitudes of the sources. The horizontal dotted line indicates $|b| = 30\degr$, the vertical dotted line indicates the weighted mean (= 0.171$\pm$0.014$\pm$0.083~Jy) of the 100\,GHz sources with significant flux density excess for sources at $|b| > 30\degr$. The true weighted mean when
including the sources without an excess is a factor of 4 lower and is between 30 and 60 mJy (see text for details). \label{fig:glat}}
\end{figure}

In Figure \ref{fig:glat}, we plot the Galactic-latitude distribution of the excess in our sample of sources with positive flux density excess. It appears that there is a plateau in the 100\,GHz excess, i.e., flux density excess is not varying with increasing Galactic latitudes. This plateau is likely the result of Eddington bias because we have not plotted any sources with negative flux density excesses. We recall that 
among the 147 high Galactic latitude ($|b|>30\degr$) sources for which the multi-band photometry
was adequate to assess the existence of CO emission, only 53 have $S_{\rm excess} > \sigma_{\rm excess}$
and 10 of them have $S_{\rm excess} > 3\sigma_{\rm excess}$. It is these 53 sources which make it appear
that there is a plateau. When including the remaining sources, the median excess at 100 GHz for all the 147 high latitude sources is 55$\pm$10 mJy, which is 4.4\% of the 100 GHz flux density of the sources and cannot be accounted for by beam uncertainties or calibration which are less than one percent.

%We estimate the plateau value by calculating the weighted mean of $S_{100}^{excess}$ for all sources at $|b| > 30\degr$. A $3\sigma$ clipping is applied to exclude eight outliers before computing the mean. We find a value of $0.171\pm0.014\pm0.083$\,Jy, which corresponds to a CO intensity of 2.7 K\,km\,s$^{-1}$ based on our fit. 

We attempt to assess the robustness of this value to systematic errors. We first undertake a Monte-carlo simulation
where we scatter the flux densities of the 147 sources at 70, 100 and 143 GHz by their photometric uncertainty and repeat the fitting process
outlined as Approach II. We find that the median excess at 100 GHz from the 1000 Monte-carlo runs resulted in 60$\pm$12 mJy with the full range of values
spanning 24 to 96 mJy. 

We also test random subsets of the sources for the excess. For example, Eddington bias affects fainter sources more than the brighter one.
If the median was measured only for the SNR$>$8 sources at 100 GHz, we would derive a median excess of 57$\pm$10 mJy. When only sources with SNR$<$30 are selected, we would derive a median excess of 30$\pm$10 mJy (See Fig. \ref{fig:xsstats}).

Thus, it appears that the excess we detect due to CO at high Galactic latitude is robust at the $3-5\sigma$ level
and equivalent to a flux density at 100 GHz between 30 and 60 mJy. Using Figure 9, we find that this would translate to a CO intensity of
0.5$\pm$0.1\,K\,km\,s$^{-1}$. 

A similarly low value has been seen in individual lines of sight, away from sources,
in the component-separated \textit{Planck} maps. Fig.~17 of \citet{planck2013-p03a}
shows evidence for low CO line intensities between 0.5 and 3\,K\,km\,s$^{-1}$ which
correlate well with ground-based measurements of CO along individual lines of sight by \citet{Hartmann98} and \citet{Magnani00}.
It is however challenging to probe the low levels of intensity we are probing here since the noise in the \citet{dame2001} CO map
is about 1\,K\,km\,s$^{-1}$ while the noise in the \citet{Hartmann98} maps is about 0.2\,K\,km\,s$^{-1}$. Substructure at the level
that we are probing would be undetected in those surveys. We also find that the Northern hemisphere appears to show stronger
emission than the Southern Hemisphere by a factor of $\sim2-3$ (2.5$\sigma$) but given the weakness of the signal in the entire catalogue, we
suggest that using deeper catalogues or deeper, higher resolution surveys from the ground are required to confirm this tentative result. If true,
a comparison with the results in \citet{Hartmann98} and \citet{Magnani00}
would suggest that the Northern Hemisphere contains fewer large clouds and more small clouds while the Southern Hemisphere
contains relatively fewer small clouds and more large clouds where the distinction between large clouds and small clouds is $\sim$15\,M$_{\sun}$. 
The most likely origin for this is likely to be the location of the Sun above the Galactic mid plane which provides an unimpeded sight line
through the lower mass clumps at high latitude. 

Finally, we assess the gas mass in the clumps that we are detecting. The filling factor of this gas is unclear since we only trace this
low level of emission from the 147 sources with high SNR. The lower limit to the filling
factor is the fractional solid angle subtended by these 100 GHz \textit{Planck} sources and is 6$\times$10$^{-4}$. Adopting the
same distance parameters and conversion factors of \citet{Hartmann98} results in a total mass
of 2060\,M$_{\sun}$ which is comparable to the mass (and mass surface density) in the large clouds at high latitude. 

\begin{figure*}
\centering
\includegraphics[width=0.45\textwidth]{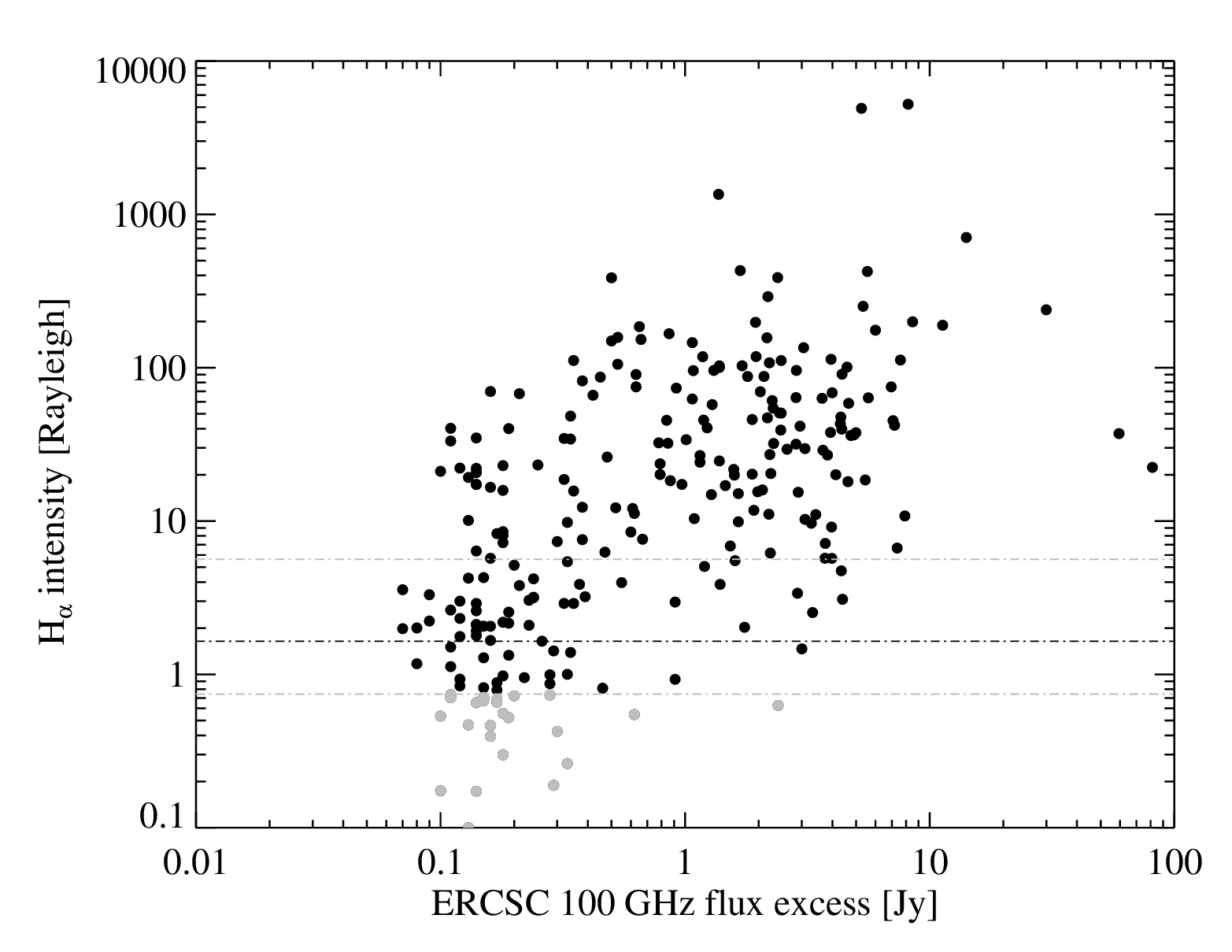} 
\includegraphics[width=0.49\textwidth]{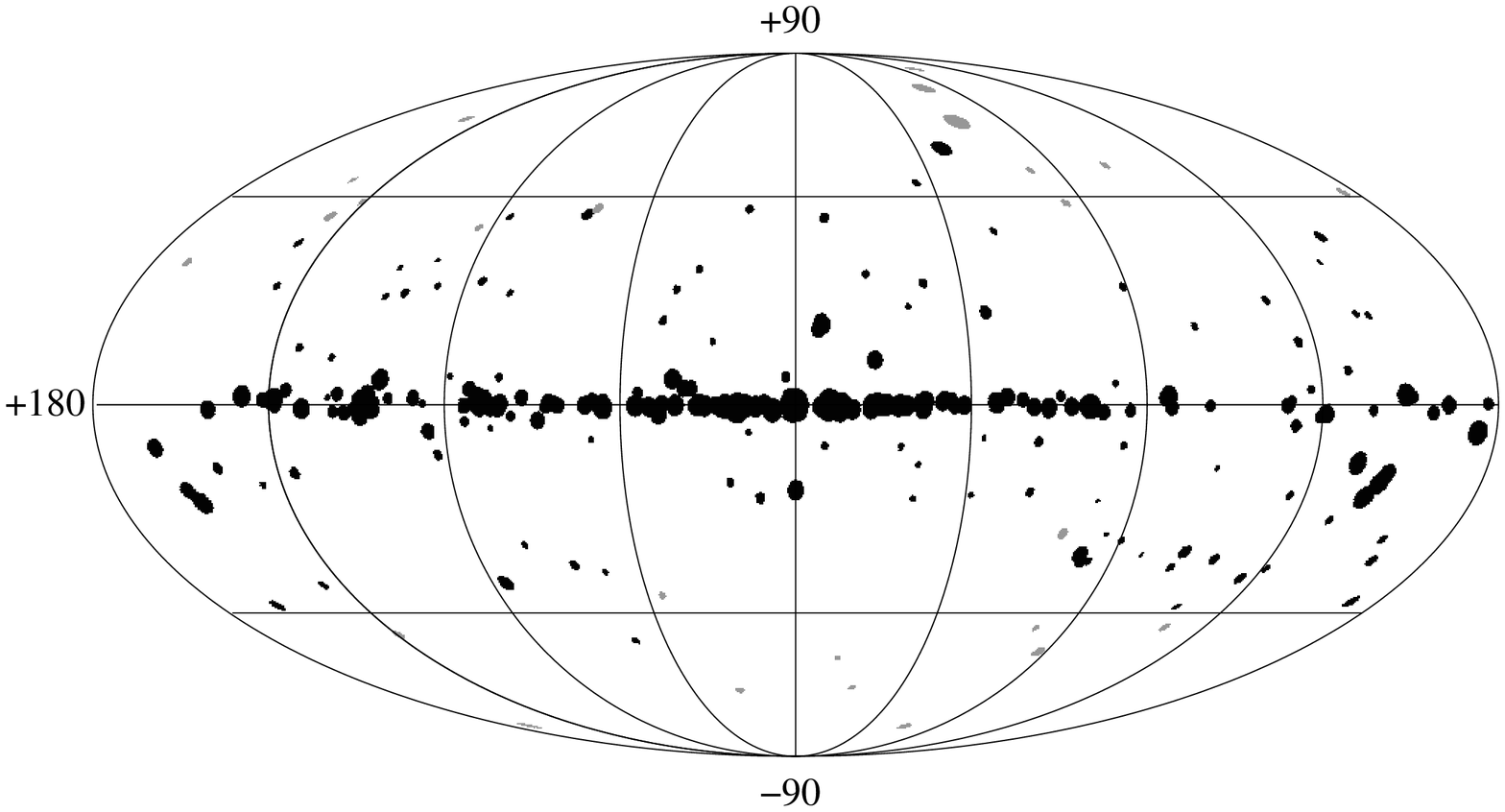} 
\caption{Left: Comparison between CO excess flux and H\,$\alpha$ intensity for all 254 sources with $S_{\rm excess} > \sigma_{\rm excess}$. The black dashed line is the median value of the H\,$\alpha$ map, grey lines are the 25 and 75 percentile values. Right: Sky distribution of ERCSC sources that have flux excess at 100\,GHz. In both plots, sources that are correlated with high H\,$\alpha$ emission are in black filled circles, while grey circles are sources with H\,$\alpha$ intensity values less than the 25 percentile of the full sky values.\label{fig:halpha}}
\end{figure*}

\begin{figure*}
\centering
\includegraphics[width=0.47\textwidth]{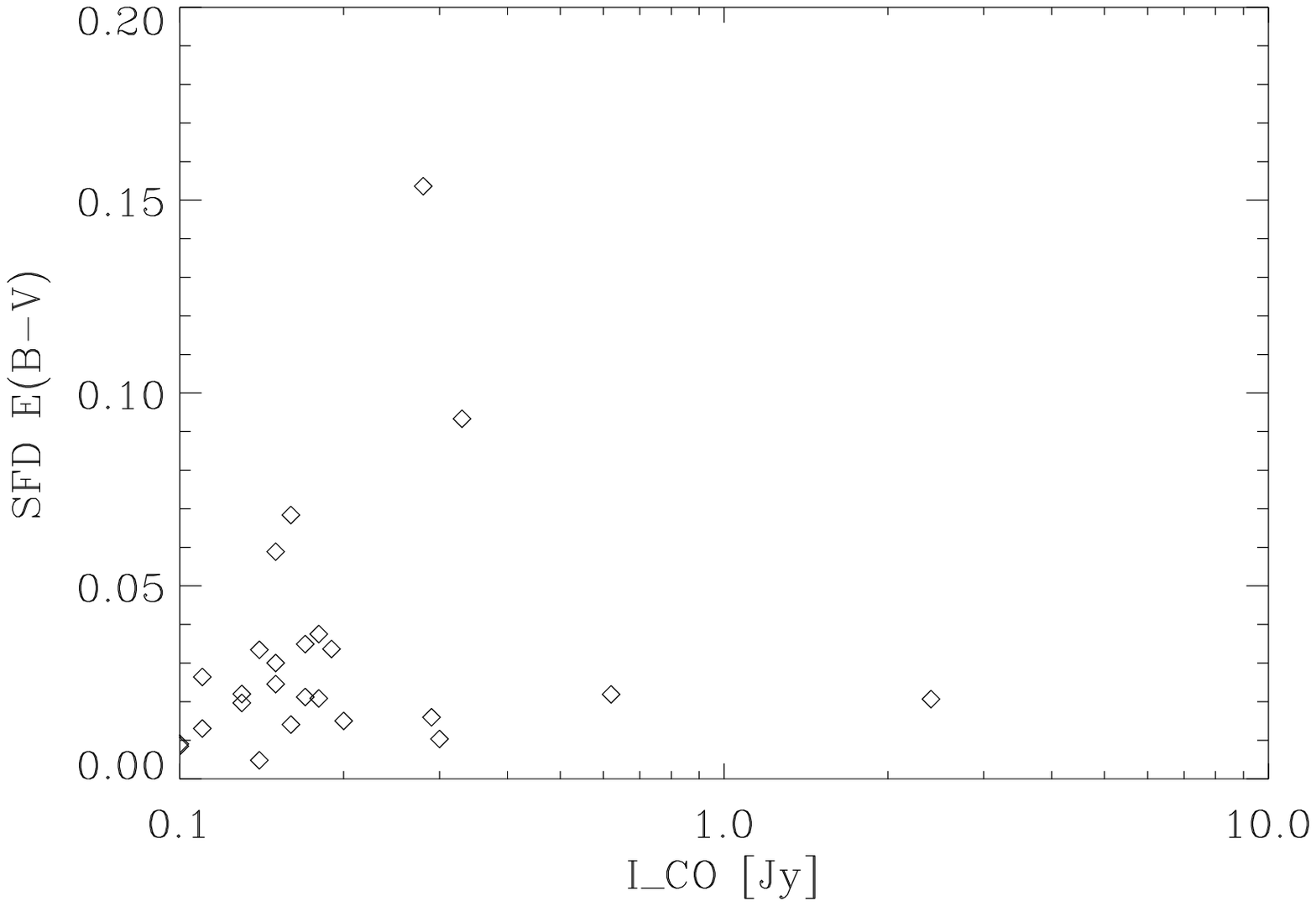} 
\includegraphics[width=0.47\textwidth]{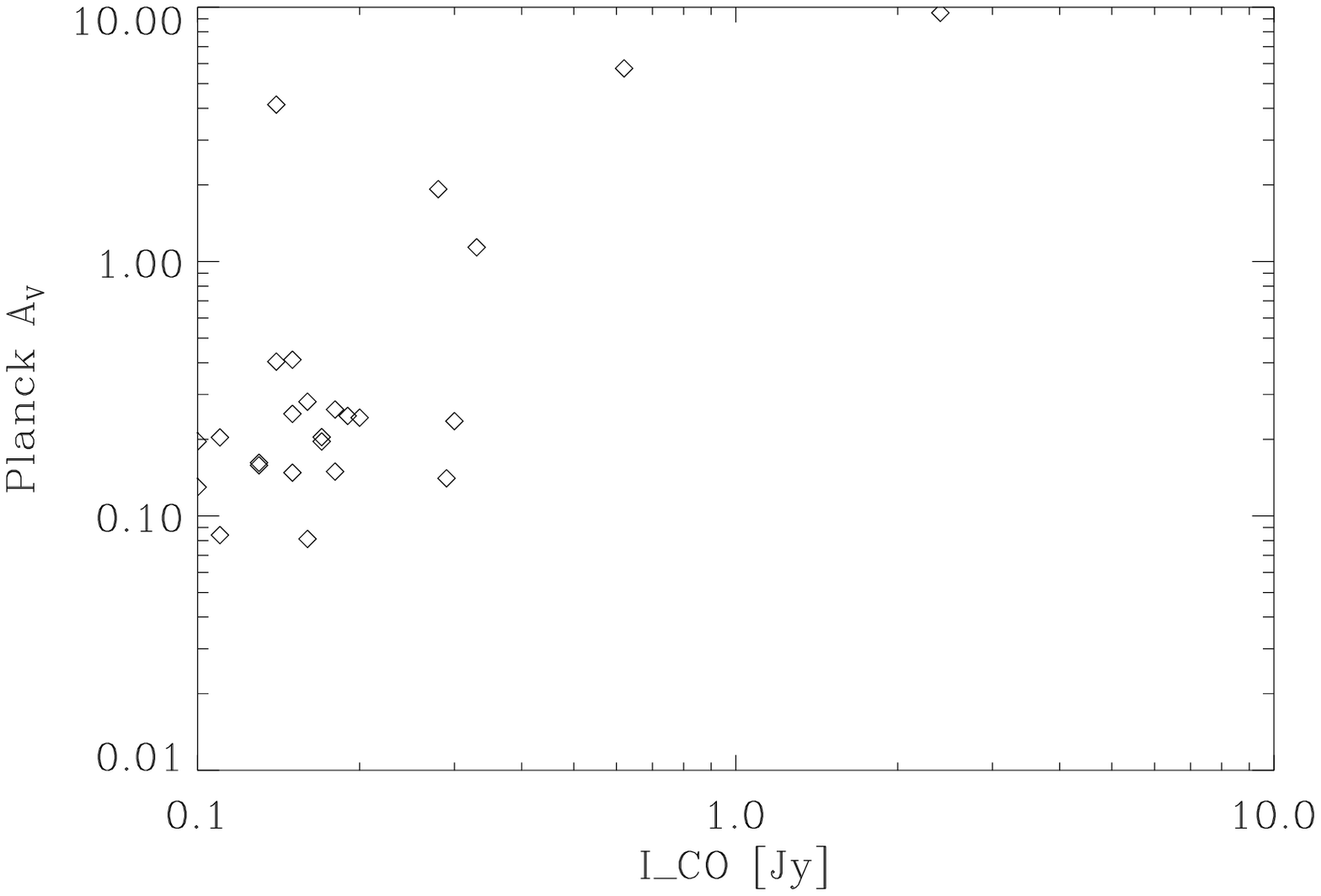} 
\caption{Comparison between our derived flux-density excess due to CO and Galactic extinction values along the line of sight for the 21 sources with low H\,$\alpha$ intensity. The fact that the extinction values are non-zero strengthens the case for there being gas associated with the
dust which is responsible for the diffuse extinction.
\label{fig:extinction}}
\end{figure*}

\subsection{Alternative explanation to the ERCSC 100\,GHz flux excess?}
Electric dipole emission from spinning dust is generally thought to contribute to the Galactic microwave emission detected between $\sim 10$ and 60\,GHz. Theoretical models suggest that the peak frequency of spinning dust emission could be as high as $\sim150$\,GHz, depending on the properties of the dust grains and their environment \citep[e.g.,][]{Draine1998, Ali2009, Ysard2010a}.  Therefore it could potentially contribute to the 100\,GHz flux-density excess in the ERCSC sources. However, such high-frequency peaks require extreme environments: very high density, moderately-high radiation field, and high fraction of ions. To date, spinning dust emission has been detected mostly at 20--40\,GHz across the Galactic sky. There has been one extragalactic detection of spinning dust emission peaked at 33\,GHz in a star-forming region in the galaxy NGC\,6946 \citep{Murphy2010}. Combining the \textit{Planck} ERCSC, \textit{WMAP} and other archival measurements, \cite{Peel2011} studied the continuum spectra of three nearby dusty star-forming galaxies (M82, NGC\,253, and NGC\,4945) but found no convincing evidence for spinning dust. 

In addition, since CO is dense in star-forming regions which can also be traced by their H\,$\alpha$ emission, we compare in Figure~\ref{fig:halpha} the 100\,GHz flux-density excess of our sample with the H\,$\alpha$ intensity of these regions in the composite H\,$\alpha$ full-sky map \citep{Finkbeiner2003}. The dark filled circles are sources with H\,$\alpha$ intensity greater than the 25 percentile value of the full-sky map and are likely to be associated with star formation regions. The relationship with H\,$\alpha$ shows a larger scatter than the correlations in Figure 9 and 10, suggesting a less direct relationship with a star-formation tracer. The formal Pearson's correlation coefficient of the excess against H$\alpha$ is 0.6 while that against the Dame CO map is 0.97.
This suggests that although some star-formation may be present especially along the lines of sight to 100\,GHz-excess sources in the Galactic plane, the CO emission that we detect
is more likely attributable to diffuse ISM and is not associated with individual star-forming regions.
There are 21 sources (grey filled circles) with H\,$\alpha$ intensity below the 25 percentile value. For these sources, we compare the flux-density excesses with measures of dust extinction, such as the \cite{Schlegel1998} $ E(B-V)$ values and the \textit{Planck} $A_V$ values (Figure~\ref{fig:extinction}). Although no strong correlation is seen, the plots show that the value of $A_V$ is not zero along those lines of sight; there is therefore dust present in these locations, which in turn is expected to be collocated with ISM gas. The formal correlation coefficient with extinction measures is 0.87.

\subsection{Excess in other bands: A consistency check}
A detection of the $^{12}$CO $J$=1$\rightarrow$0 emission in the 100 GHz band motivates a check
for the next strongest line, the CO $J$=2$\rightarrow$1 emission in the 217 GHz bands. 
Indeed, during component separation of the diffuse foregrounds, \cite{planck2013-p03a} find evidence for significant contribution of CO $J$=2$\rightarrow$1 
emission to the 217 GHz intensity and CO $J$=3$\rightarrow$2 emission to the 353 GHz intensity, albeit mainly in the Galactic plane. To be able to undertake the same analysis,
at high latitude for the $J$=2$\rightarrow$1 line, a source needs to be detected at 143, 217 and 353 GHz in
addition to 70 and 100 GHz. The 79 ERCSC sources seen in all bands are 
a robust, bright sample for this analysis. Reproducing the analysis 
that was performed to derive the 100 GHz excess at the 143, 217 and 353 GHz,
we find that of the 69 sources that have an excess at 100 GHz, only 14 have
an excess at 217 GHz and only 2 having a statistically significant
excess ($S_{\rm excess} > 3\sigma_{\rm excess}$) at both 100 and 217 GHz. This is partly because
sources are typically brighter at 217 GHz due to the rising thermal dust emission and the
fractional increase in their brightness due to CO $J$=2$\rightarrow$1 is a smaller fraction
of their flux density and typically well within their flux density uncertainty. 
In addition, the bandwidth at 217 GHz is broader. Thus for the same CO line flux, the 
broadband flux density excess is half of what it is at 100 GHz, although that
is compensated for in a limited way by the reduced noise in the 217 GHz map.
For the two sources which do have a statistically significant excess
at both 100 and 217 GHz, PLCKERC G250.08-31.09 and PLCKERC G305.11+57.05,
they are both at high latitude and with 2$\rightarrow$1 /1$\rightarrow$0  ratios of 1.4 and 1.1
respectively, consistent with what is seen in the diffuse ISM where the ratio ranges from 0.5 to 1.5 and in individual galaxies
where it may be as high as 5.
For the 55 sources which do not show a 217 GHz excess, the median upper limit of 2$\rightarrow$1 /1$\rightarrow$0 is 0.7 further
indicating that the 100 GHz excess is robust.
We do not find any high latitude sources which have possible excess at 353 GHz due to CO 3$\rightarrow$2 emission
and also at 100 GHz.

We then attempt the converse; which is we attempt to measure an excess at 217 and 353 GHz from CO 2$\rightarrow$1
and CO 3$\rightarrow$2 emission, using the same methodology adopted earlier. We then assess if the excess in those
bands is consistent with an absence of excess in the 100 GHz band. We note that this calculation is fraught with uncertainty
because at these higher frequencies, particularly 353 GHz, the contamination from the Rayleigh-Jeans tail of cold dust emission is significant (Figure 5.).
There are 95 sources for which there may be an excess at 217 GHz; comparing these regions with the Dame CO map shows a weak correlation
with a correlation coefficient of 0.6. 
For the 4 sources at high latitude which have an apparent excess at 217 GHz, we find that two have 
lower limits of 2$\rightarrow$1 /1$\rightarrow$0 of 1.4 and 1.1, consistent with the range seen in the ISM. One other source is strongly contaminated
by rising dust emission even at 217 GHz while the fourth source has a lower limit of 4, which is unusual but consistent with the range seen.

Repeating the analysis at 353 GHz, we find that of the 287 high latitude sources which have an apparent excess, seven have any observational constraints at 100 GHz.
Four have low significance
excess at 100 GHz while the rest have no evidence of excess at 100 GHz. If this excess were due to CO 3$\rightarrow$2, the median lower limit would be $\sim$50 which is unphysical
since values in star-forming galaxies are typically 5-10. 
For the one source which has excess from both CO 3$\rightarrow$2 and 2$\rightarrow$1 (but not constraints at 100 GHz), the intensity ratio is 5.7, at the high end but consistent with values seen in star-forming galaxies.
As a result, we conclude that although the methodology we have adopted works well up to 217 GHz, at higher frequencies where dust properties
show a wide range of far-infrared color temperatures, it is challenging to use an interpolation between adjacent bands to find evidence for CO 3$\rightarrow$2 excess.

Thus, although a useful consistency check, we find that we do not have
either the S/N or source statistics to claim a robust high latitude excess at the other 
bands although the average excess, where detectable, is consistent with the line ratios seen in \cite{planck2013-p03a}.

\subsection{Contamination from nearby sources at 143 GHz}
Another possible contribution to the excess at 100 GHz is contamination
of the 100 GHz flux density from a faint nearby source. Since the 100 
GHz beam is larger than that of the 143 GHz band, an association of 
only one of multiple nearby 143 GHz sources to a 100 GHz source would result in an excess being artificially estimated at 100 GHz.
To assess this, we have cross-matched the list of excess sources with the 
deeper, low-SNR 143 GHz catalogue from Planck Data Release 1 which is also
available on the public Planck archive. We find that only 2 of the 
53 excess sources at high latitude have two close 143 GHz counterparts within 
1.5$\times$FWHM at 100 GHz and only 1 of the excess sources has the counterpart within 
1 FWHM. This rules out the presence of contamination from faint nearby sources playing a significant role especially
at high Galactic latitude. Naturally, there can be 
contamination close to the Galactic plane. In fact, we find that out of the 255 sources, 84 have a close companion within 1.5$\times$FWHM. However, our primary result is for the contribution
of CO in the high latitude sky and hence the contamination in the Plane does
not affect our conclusions.

\begin{figure*}
\centering
\includegraphics[width=0.47\textwidth]{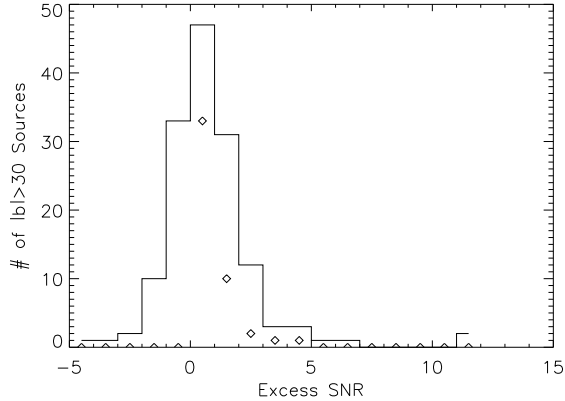} 
\caption{The solid line is the histogram of S/N ratio
of the excess (both ``positive" and ``negative") at 100 GHz for all high
latitude sources, calculated as described in the paper. The diamonds
are the negative side of the histogram flipped on the positive side, i.e. if the
excess was pure noise with a mean of zero, the distribution would look like the diamonds
at the positive values. Clearly, the number of positive excess sources
is above what would be estimated for pure noise. Similar plots generated
at other frequencies are noise dominated.
\label{fig:xsstats}}
\end{figure*}

\section{Conclusions}
\label{sec:conclusion}

We have created a bandmerged catalogue of the \textit{Planck} Early Release Compact Source Catalog in this paper, and we have shown how confusing cases, arising from the varying spatial resolution in the different bandpasses, can be handled in the process. We used the catalogue to probe the spectral types of the sources at different \textit{Planck} frequencies. The flux densities in the \textit{Planck} 100\,GHz bandpass are contaminated by CO $J$=1$\rightarrow$0 emission. We have shown that by simple interpolation between the 70 and 143\,GHz flux densities of individual sources, we can estimate the flux-density excess arising from the CO along the lines of sight. The CO excess agrees well with the intensity from the observed CO map in regions close to the Galactic plane, and is also correlated with the IRAS 100$\micron$ sky background, which is dominated by dust emission.
Thus, we are able to use the multi-wavelength spectra of bright
sources to probe the CO distribution at high Galactic latitudes where the first estimates from ground-based observations and \textit{Planck}
have only recently become available
\citep{planck2013-p03a, Magnani00, Hartmann98}.
We find evidence that the high latitude CO emission is structured on 10$\arcmin$ scales and contributes a median line of sight intensity of 0.5$\pm$0.1\,K\,km\,s$^{-1}$, 
which we argue is evidence for a population of low mass ($\sim15\,M_{\sun}$) high latitude, molecular gas clumps. There is $\sim2.5\sigma$ evidence for
asymmetricity between the Northern and Southern Galactic hemispheres where the Northern hemisphere seems to show more small clumps than the Southern hemisphere. 
These clumps could account for a lower limit of 2000\,M$_{\sun}$ of molecular gas at high latitude. Deeper, higher resolution CO surveys will be required to provide an
accurate census of these clumps and their contribution to the total molecular gas budget.

%%%%%%%%%%%%%%%%%%%%%%%%%%%%%%%%%%%%%%%%%%%%%%%%%%%%%%%%%%%%%%%%%%%%%%%%%%%%%%%%%

\section*{acknowledgements}
We thank the referee for thoughtful comments which improved the manuscript. We also thank Dr. Clive Dickinson for helpful discussions in the preparation of this paper. This research has made use of the NASA/IPAC Extragalactic Database (NED) and NASA/ IPAC Infrared Science Archive, which are operated by the Jet Propulsion Laboratory, California Institute of Technology, under contract with the National Aeronautics and Space Administration. Some of the results in this paper have been derived using the HEALPix \citep{gorski2005} package.

\clearpage

\bibliographystyle{mn2e}
\newcommand{\mnras}{MNRAS}
\newcommand{\apj}{ApJ}
\newcommand{\apjl}{ApJL}
\newcommand{\apjs}{ApJS}
\newcommand{\aap}{A\&A}

\bibliography{Planck_bib,xi_refs}

\end{document}